\UseRawInputEncoding
\documentclass[reprint,
superscriptaddress,
amsmath,amssymb,
aps,pra,prb,rmp,prstab,prstper,floatfix]{revtex4-2}
\usepackage[dvipsnames]{xcolor}
\usepackage{graphicx}% Include figure files
\usepackage{dcolumn}% Align table columns on decimal point
\usepackage{bm}% bold math
\usepackage[mathlines]{lineno}% Enable numbering of text and display math
\usepackage{hyperref}
\hypersetup{colorlinks=true,linkcolor=black,citecolor=black}
\usepackage{graphicx}
\usepackage{multibib}
\usepackage{makecell}
\usepackage{array}

\begin{document}

\title{ Dynamical topology of chiral and nonreciprocal state transfers in a non-Hermitian quantum system}
\author
{Pengfei Lu$^{1,2,4}$, Yang Liu$^{1,2,3,4}$, Qifeng Lao$^{1}$, Teng Liu$^{1}$, Xinxin Rao$^{1}$, Ji Bian$^{1}$, \\Hao Wu$^{1}$, Feng Zhu$^{1,2}$, Le Luo$^{1,2,3,\ast}$\\
	\normalsize{$^{1}$School of Physics and Astronomy, Sun Yat-sen University, Zhuhai 519082, China.}\\
	\normalsize{$^{2}$Center of Quantum Information Technology,}\\
	\normalsize{Shenzhen Research Institute of Sun Yat-Sen University, Shenzhen 518087, China.}\\
	\normalsize{$^{3}$Quantum Science Center of Guangdong-HongKong-Macao Greater Bay Area, }\\
	\normalsize{Shenzhen 518048, China.}\\
	\normalsize{$^{4}$These authors contributed equally to this work.}\\
	\normalsize{$^\ast$Corresponding author. E-mail: luole5@mail.sysu.edu.cn}
}

%\date{\today}

\begin{abstract}
The fundamental concept underlying topological phenomena posits the geometric phase associated with eigenstates. In contrast to this prevailing notion, theoretical studies on time-varying Hamiltonians allow for a new type of topological phenomenon, known as topological dynamics, where the evolution process allows a hidden topological invariant associated with continuous flows. To validate this conjecture, we study topological chiral and nonreciprocal dynamics by encircling the exceptional points (EPs) of non-Hermitian Hamiltonians in a trapped ion system. These dynamics are topologically robust against external perturbations even in the presence dissipation-induced nonadiabatic processes. Our findings indicate that they are protected by dynamical vorticity---an emerging topological invariant associated with the energy dispersion of non-Hermitian band structures in a parallel transported eigenbasis. The symmetry breaking and other key features of topological dynamics are directly observed through quantum state tomography. Our results mark a significant step towards exploring topological properties of open quantum systems.
\end{abstract}

\maketitle

\section*{Introduction}
Geometric phases play a crucial role in classifying gapped quantum systems that are protected by the symmetry of Hamiltonians~\cite{schnyder2008classification,chiu2016classification}, often resulting in robust physical properties that are resilient to perturbations. In time-varying cases, if the evolution is sufficiently slow, the system remains in the eigenstate of the instantaneous Hamiltonian, thereby preserving the geometric phases during dynamic processes.
This phenomenon, validated in various Hermitian systems~\cite{cohen2019geometric,zhang2005experimental,hasan2010colloquium,xu2011topological}, relies on the preservation of symmetry and adherence to the adiabatic condition. One might expect that this similarly applies to time-varying non-Hermitian systems; However, this expectation does not hold. When encircling the exceptional points (EPs) in a non-Hermitian system, the state deviates from the eigenstate of the instantaneous Hamiltonian regardless of how slowly the system evolves~\cite{doppler2016dynamically,xu2016topological}. This deviation is known as dissipation-induced nonadiabatic transitions (DNATs). The occurrence of DNATs disrupts the continuous accumulation of geometric phases on the Riemann surfaces of complex eigenvalues. Consequently, characterizing the topological properties of dynamical non-Hermitian systems remains a significant and unresolved question.

Previous research on encircling EPs in non-Hermitian systems reveals fascinating topological behaviors, which have been experimentally explored in both classical and quantum systems, including microwave/optical setups~\cite{dembowski2001experimental,doppler2016dynamically,zhang2018dynamically,yoon2018time,li2020hamiltonian,nasari2022observation,yang2023realization}, optomechanical oscillators~\cite{xu2016topological}, acoustic cavities~\cite{ding2016emergence,tang2020exceptional}, superconducting circuits~\cite{abbasi2022topological}, NV centers~\cite{liu2021dynamically}, and cold atoms~\cite{ren2022chiral}.
Some of these experiments have demonstrated that in the absence of DNATs, a parameter variation encircling an EP causes two states to switch positions after one cycle and acquires a geometric phase of $\pi$ after two cycles. With DNATs, however, the geometric phase becomes elusive during dynamic processes. Instead, the vorticity of energy eigenvalues may serve as a new type of topological invariant in non-Hermitian systems~\cite{shen2018topological,kawabata2018anomalous}, but its validity depends on the evolution trajectories remaining on the Riemann surfaces, which is contradicted by the presence of DNATs.
The most intriguing topological behaviors in these systems~\cite{doppler2016dynamically,xu2016topological,zhang2018dynamically}, such as chiral state transfers and nonreciprocal mode switching, depend on the occurrence of DNATs. Therefore, it is essential to uncover a profound relationship between DNATs and hidden topological invariants in dynamics. This understanding could help identify the topological structure of non-Hermitian dynamics.

In this work, we experimentally study the topological chiral and nonreciprocal state transfers by dynamically encircling the EPs of both $\mathcal{PT}$ (parity-time) ($[H, \mathcal{PT}] = 0$) and $\mathcal{APT}$ (anti-parity-time) symmetric ($\{H, \mathcal{PT}\}=0$) Hamiltonians in a trapped-ion system. The outcomes unveil a universal rule for determining the behavior of state transfers associated with chiral and time-reversal symmetries. Chiral state transfers, where encircling an EP in a clockwise or counterclockwise direction results in different final states, are exclusively dictated by the symmetry of the initial effective Hamiltonian in the parallel transported eigenbasis. Nonreciprocal state transfers, where the processes depend on the initial states and the direction of encircling the EP, are dictated by the time derivative of the initial states. We find that these state transfers are robust against substantial external noise introduced during the encircling processes. To reveal the underlying topological mechanism of this robustness, we transform the original Hamiltonian into the case in the parallel transported eigenbasis and define a new topological invariant, dynamic vorticity. This dynamic vorticity remains invariant during the encircling process, regardless of whether dissipation-induced nonadiabatic transitions (DNATs) occur, and is solely determined by the number of EPs that the trajectories encircle. This reveals that both chiral and nonreciprocal state transfers are protected by dynamic vorticity, which can be considered a universal topological invariant in time-varying non-Hermitian systems. These discoveries validate that topological dynamics can arise from the interplay of dissipation and coherence, opening new avenues to explore the topological properties of open quantum systems.

\section*{Results}
\subsection*{Dynamically encricling the EPs of $\mathcal{PT}$ and $\mathcal{APT}$ symmetric Hamiltonians}
\begin{figure*}
	\centering
	\includegraphics[width=0.9\textwidth]{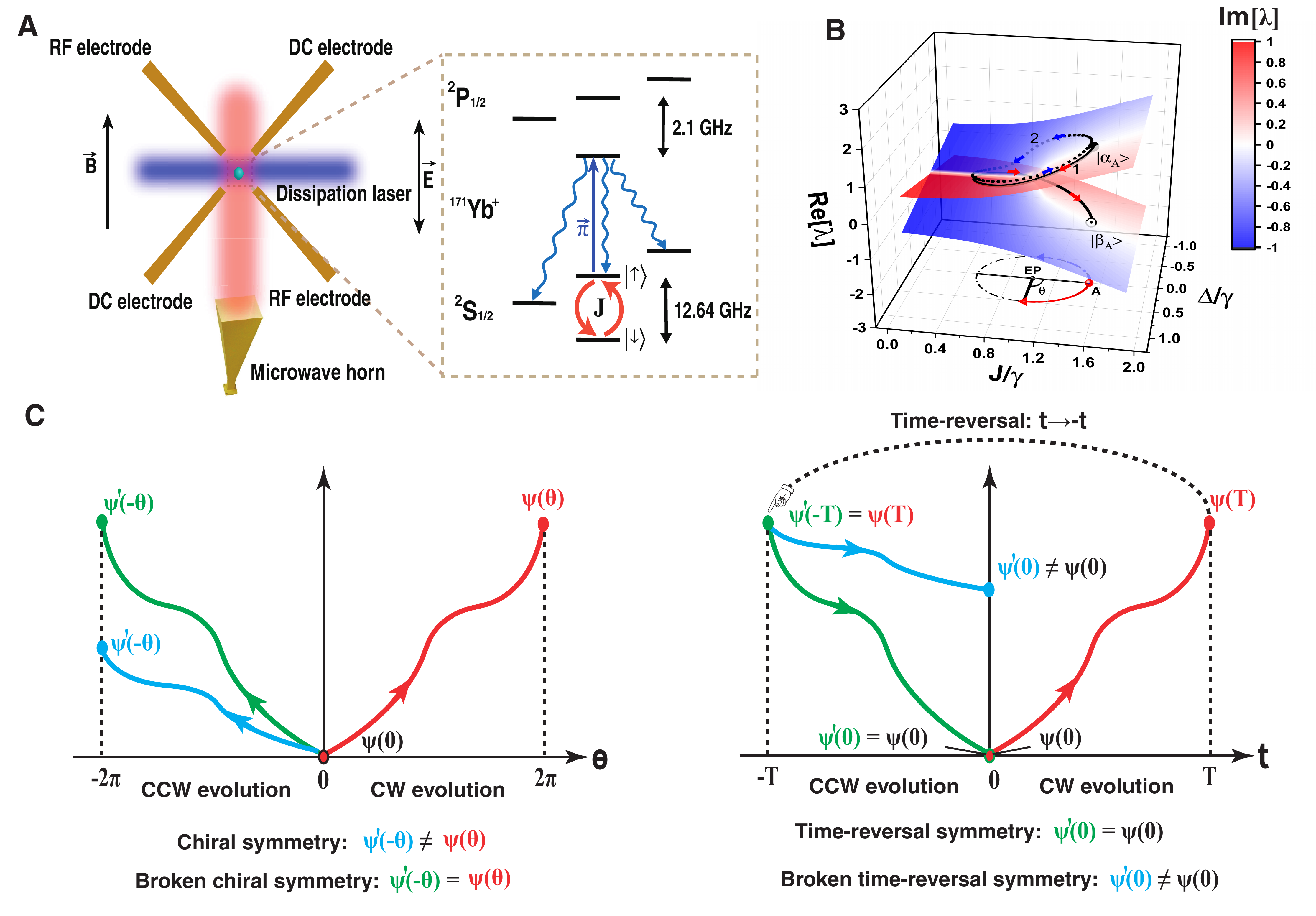}
	\caption{Generation of chirality and nonreciprocity in a dissipative trapped-ion qubit. (\textbf{A})~Schematic diagram of the energy levels of a $^{171}$Yb$^{+}$ ion. The five involved levels include $|F=0, m_F =0\rangle$ and $|F=1, m_F =0, \pm1\rangle$ in the electronic ground state $^2 S_{1/2}$, and $|F=0, m_F =0\rangle$ in the electronic excited state $^2 P_{1/2}$. (\textbf{B})~Encircling paths in the parameter space (coupling strength $J$ and detuning $\Delta$) and the state evolution trajectories projected onto the eigenvalues' Riemann sheets, starting in the $\mathcal{PT}$-symmetric ($\mathcal{PTS}$) regime of the $\mathcal{PT}$ Hamiltonian. The solid (dashed) trajectory in (B) denotes the clockwise (counterclockwise) evolution of $|\alpha_A(0)\rangle$. (\textbf{C})~Chiral symmetry (depicted by red and green curves in the left panel) and broken chiral symmetry (depicted by red and blue curves in the left panel) serve as mathematical criteria for determining chiral dynamics. Time-reversal symmetry (illustrated by red and green curves in the right panel) and broken time-reversal symmetry (illustrated by red and blue curves in the right panel) serve as mathematical criteria for determining nonreciprocal dynamics.}
	\label{encircling_setup}
\end{figure*}

The experiments utilize a passive $\mathcal{PT}$-symmetric system involving a single trapped $^{171}$Yb$^+$ ion, building upon the methodologies detailed in our previous works  ~\cite{lu2024realizing,bian2023quantum,bian2022PhRvA}. The ion is confined and laser-cooled in a linear Paul trap. Then, it is initialized to the hyperfine state $\mid\downarrow\rangle=|F = 0, m_F = 0, { }^2 S_{1 / 2}\rangle$ of the ground state by optical pumping. The qubit levels, hyperfine states $\mid\downarrow\rangle$ and $\mid\uparrow\rangle=|F = 1, m_F= 0, { }^2 S_{1 / 2}\rangle$, are driven by a microwave with a coupling rate $J$. The dissipation is introduced by resonantly driving a transition from $\mid\uparrow\rangle$ to $|F = 1, m_{F} = 0, {}^{2}P_{1/2}\rangle$ of the excited state, resulting in the spontaneous decay to three ground states $|F = 1, m_{F} = 0, \pm 1\rangle$ in $^{2}S_{1/2}$ with equal probability. Taken the Zeeman sublevels $|F = 1, m_{F} = \pm 1\rangle$ in $^{2}S_{1/2}$ as an auxiliary state $|a\rangle$, the system is equivalent to a spin-dependent loss from the qubit state $\mid\uparrow\rangle$ to state $|a\rangle$ with a rate at $4\gamma$, as the coupling of the qubit to the environment. When the coupling strength of the dissipation beam is significantly lower than the linewidth of the $^2 P_{1/2}$ excited state, the involved energy levels can be approximated to a dissipative two-level system, as illustrated in Fig.~\ref{encircling_setup}A. Consequently, a passive $\mathcal{PT}$-symmetric non-Hermitian Hamiltonian is derived ~\cite{li2019observation, lu2024realizing,naghiloo2019quantum}, $H_{eff}=J\hat{\sigma}_x+i\gamma\hat{\sigma}_z-i\gamma\hat{I}$, where $\hat{\sigma}_x=\mid\downarrow\rangle\langle\uparrow\mid+\mid\uparrow\rangle\langle\downarrow\mid$, $\hat{\sigma}_z=\mid\downarrow\rangle\langle\downarrow\mid-\mid\uparrow\rangle\langle\uparrow\mid$, and $\hat{I}=\mid\downarrow\rangle\langle\downarrow\mid+\mid\uparrow\rangle\langle\uparrow\mid$,
which can be mapped to a $\mathcal{PT}$-symmetric Hamiltonian by adding an extra term $i\gamma\hat{I}$.

To encircle the EP of the $\mathcal{PT}$ symmetric Hamiltonian, the driving microwave features time-varying detuning and intensity, resulting in a time-dependent Hamiltonian expressed as ($\hbar=1$)
\begin{equation}
	H(t) = \begin{pmatrix} \Delta(t)/2+i \gamma & J(t) \\ J(t) & -\Delta(t)/2-i \gamma \end{pmatrix},
	\label{Heff}
\end{equation}
where $\Delta(t)$ and $J(t)$ represent the detuning and the coupling rate of the microwave, respectively, and $\gamma$ denotes the dissipation rate dependent on the laser intensity. The eigenvalues of Eq.~(\ref{Heff}) are complex numbers given by  $\lambda_{1,2}(t)=\pm \sqrt{J(t)^2-\gamma(t)^2+i \gamma(t) \Delta(t)+\Delta(t)^2 / 4}$, with the EP occurring at $\Delta=0$ and $J=\gamma$.

The system evolves by varying $\Delta(t)$ and $J(t)$, while keeping $\gamma$ fixed. The real and imaginary components of $\lambda_{1,2}(t)$ form two intersecting Riemann sheets wrapped around the EP, as shown in Fig.~\ref{encircling_setup}B. We drive the qubit through a parameter loop defined by $\Delta(t)= r \sin[\theta (t) +\theta_0] $ and $J(t)= J_0 + r \cos[\theta(t) +\theta_0]$, where $r$ represents the encircling radius, $\theta(t)$ denotes the time-dependent encircling angle, and $\theta_0=0$ ($\theta_0=\pi$) determines the starting point in the $\mathcal{PT}$-symmetric (broken) regime, abbreviated as the $\mathcal{PTS}$ ($\mathcal{PTB}$) regime. More details regarding the encircling procedure can be found in Materials and Methods.

The clockwise and counterclockwise encircling trajectories from the eigenstate $\left|\alpha_A(0)\right\rangle$ ($\left|\beta_A(0)\right\rangle$) are demonstrated in Fig.~\ref{3D-PTS}, A and C (E and G), where the black lines show the state evolution projected onto the Riemann surfaces. Fig.~\ref{3D-PTS}, B and D (F and H) show the overlap between the instantaneous eigenstate $|\alpha(t)\rangle$ ($|\beta(t)\rangle$) with the evolutionary state $|\psi(t)\rangle=C_1(t)|\alpha(t)\rangle+C_2(t)|\beta(t)\rangle$, i.e. $\left\langle\alpha_A(t)|\psi(t)\right\rangle$ ($\left\langle\beta_A(t)|\psi(t)\right\rangle$). When a DNAT occurs, the qubit invariably jumps from the loss sheet (blue) to the gain sheet (red) on the Riemann surface. Such jump is characterized by the crossing between $\left\langle\alpha_A(t)|\psi(t)\right\rangle$ and $\left\langle\beta_A(t)|\psi(t)\right\rangle$ in the cyan shaded region. Its occurrence is measured through quantum state tomography of $|\psi(t)\rangle$ during these experiments.
We also investigate the dynamical encircling with $\mathcal{APT}$ symmetric Hamiltonians, constructed by sandwiching a passive $\mathcal{PT}$-symmetric Hamiltonian $H_{eff}$ between two $\pi/2$ pulses along the $\pm Y$ axis on the Bloch sphere~\cite{bian2023quantum}. The details are provided in Supplementary Material S2.

\subsection*{Chiral and nonreciprocal state transfer}
\begin{figure*}
	\centering\includegraphics[width=0.9\linewidth]{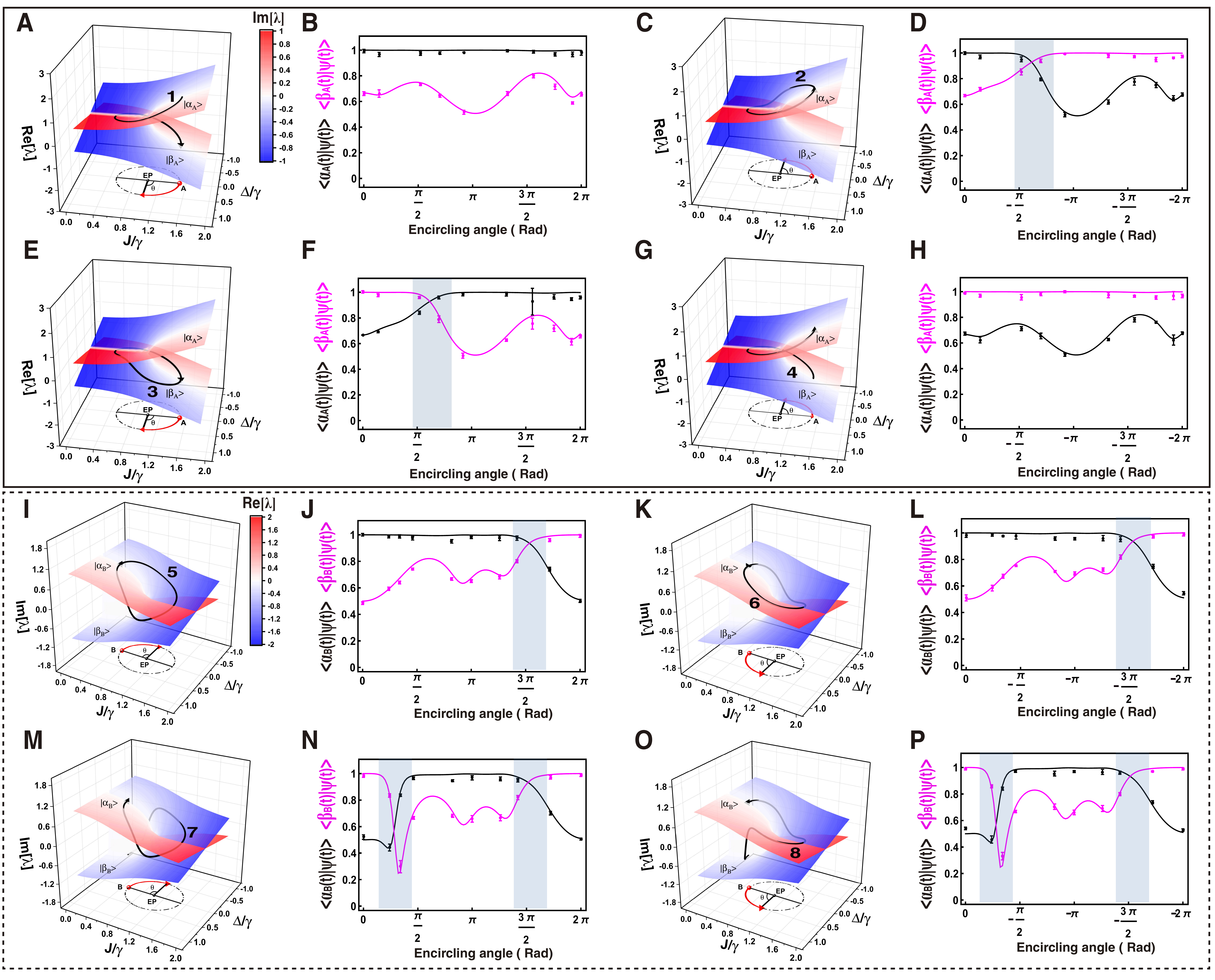}
	\caption{The time-varying evolutionary state $|\psi(t)\rangle$ starts from either the $\mathcal{PTS}$ or $\mathcal{PTB}$ regime with the $\mathcal{PT}$ Hamiltonian. The solid (dashed) box depict the state evolution trajectories in the visual four-dimensional picture of the eigenspectrum as a function of the $\Delta$ and $J$, with the colors on the Riemann sheet corresponding to the respective imaginary (real) values. Clockwise and counterclockwise encircling the EP starting from $|\alpha_{A}(0)\rangle$ ($|\alpha_{B}(0)\rangle$ ) and $|\beta_{A}(0)\rangle$ ($|\beta_{B}(0)\rangle$) are represented by trajectory 1-4 in (\textbf{A}), (\textbf{C}), (\textbf{E}) and (\textbf{G}) (trajectory 5-8 in (\textbf{I}), (\textbf{K}), (\textbf{M}) and (\textbf{O})). Figs. (\textbf{B}), (\textbf{D}), (\textbf{F}) and (\textbf{F}) ((\textbf{J}), (\textbf{L}), (\textbf{N}) and (\textbf{P})) represent the overlap between the two instantaneous eigenstates and the evolutionary state $\langle \alpha_{A}(t)|\psi(t)\rangle$, $\langle \beta_{A}(t)|\psi(t)\rangle$ ($\langle \alpha_{B}(t)|\psi(t)\rangle$, $\langle \beta_{B}(t)|\psi(t)\rangle$ ) from $\mathcal{PTS}$ ($\mathcal{PTB}$) regime, where the nonadiabatic dynamics emerge in the cyan shaded regions. Circles and lines represent experimental measurements and numerical simulation, respectively.}
	\label{3D-PTS}
\end{figure*}

The analyses of chiral behaviors in state transfer are depicted in the left panel of Fig.~\ref{encircling_setup}C. The clockwise evolution of $|\alpha_A(0)\rangle \xrightarrow{\text{CW}} |\beta_A(0)\rangle$ (trajectory 1) and the counterclockwise case of $|\alpha_A(0)\rangle \xrightarrow{\text{CCW}} |\alpha_A(0)\rangle$ (trajectory 2) exhibit the chiral state transfer, whose evolution trajectories are shown in Fig.~\ref{3D-PTS}, A to D. Another example of chiral state transfer involves the pair of of $|\beta_A(0)\rangle \xrightarrow{\text{CW}} |\beta_A(0)\rangle$  (trajectory 3) and $|\beta_A(0)\rangle \xrightarrow{\text{CCW}} |\alpha_A(0)\rangle$ (trajectory 4), as shown in Fig.~2, E to H. It is noted that the chiral state transfers are associated with starting points located in the $\mathcal{PTS}$ regime. The paths in opposite directions experience unequal numbers of DNATs, preserving the chiral symmetry. Conversely, for the starting point in the $\mathcal{PTB}$ regime, whether starting from  $|\alpha_B(0)\rangle$ (Fig.~\ref{3D-PTS}, I to L) or $|\beta_B(0)\rangle$ (Fig.~~\ref{3D-PTS}, M to P), both clockwise and counterclockwise paths experience the same number of DNATs, breaking the chiral symmetry.

The right panel of Fig.~1C explains how nonreciprocity is generated due to broken time-reversal symmetry (TRS).
Starting from the eigenstates in the $\mathcal{PTS}$ regime, the forward-time evolution (trajectory 1) and the backward-time evolution (trajectory 4) constitute a pair of reciprocal processes, the same as trajectory 5 and trajectory 6 with the initial eigenstate in the $\mathcal{PTB}$ regime. For both cases, the same number of DNATs occurs in the forward-time and backward-time paths. On the contrary, if the forward-time and backward-time encircling have different numbers of the DNATs, as seen in trajectories 3 and 4,  trajectories 2 and 1, trajectories 7 and 6, and, trajectories 8 and 5, TRS is broken, resulting in nonreciprocal state transfers.

In addition, we systematically investigate
the chirality and nonreciprocity with the $\mathcal{APT}$ Hamiltonian. The emergence of chiral (nonchiral) behavior in state transfer occurs when the starting point is in the $\mathcal{PTB}$ ($\mathcal{PTS}$) regime, opposite to the $\mathcal{PT}$-symmetric Hamiltonian. Regarding the chiral (nonreciprocal) properties, we discover a dual relationship between the $\mathcal{PT}$-Hamiltonian and the $\mathcal{APT}$ one, where the $\mathcal{PTS}$ regime of the $\mathcal{APT}$ Hamiltonian is equivalent to the $\mathcal{PTB}$ regime of the $\mathcal{PT}$ Hamiltonian, and vice versa (See Supplementary Material S2).

Based on the aforementioned observations, we ascertain that the unequal numbers of DNATs appearing in a pair of encircling paths result in chiral (nonreciprocal) state transfers.
The crucial role of DNATs stems from the dynamical phase accumulated as the parameters of the Hamiltonian evolve adiabatically~\cite{garrison1988complex,milburn2015general}. The evolutionary state is expressed as $|\psi_n(t)\rangle$ =$ e^{-\frac{i}{\hbar}\int E_n(\mathbf{R}(t'))dt'} \cdot e^{i\gamma(t)}|n(\textbf{R}(t))\rangle$, where $e^{-\frac{i}{\hbar}\int E_n(\mathbf{R}(t'))dt'}$ is dynamical phase term, $e^{i\gamma(t)}$ is geometric phase term, and $|n(\textbf{R}(t))\rangle$ is instantaneous eigenstate state of Hamiltonian. When the state evolves on the loss sheet, the dynamical phase contributes a decaying factor. Consequently, the state evolving on the loss sheet will transition to the gain sheet, whereas the state on the gain sheet will remain unaffected.

\subsection*{Classification of nonadiabatic transition}
\begin{figure*}
	\centering
	\includegraphics[width=0.8\textwidth]{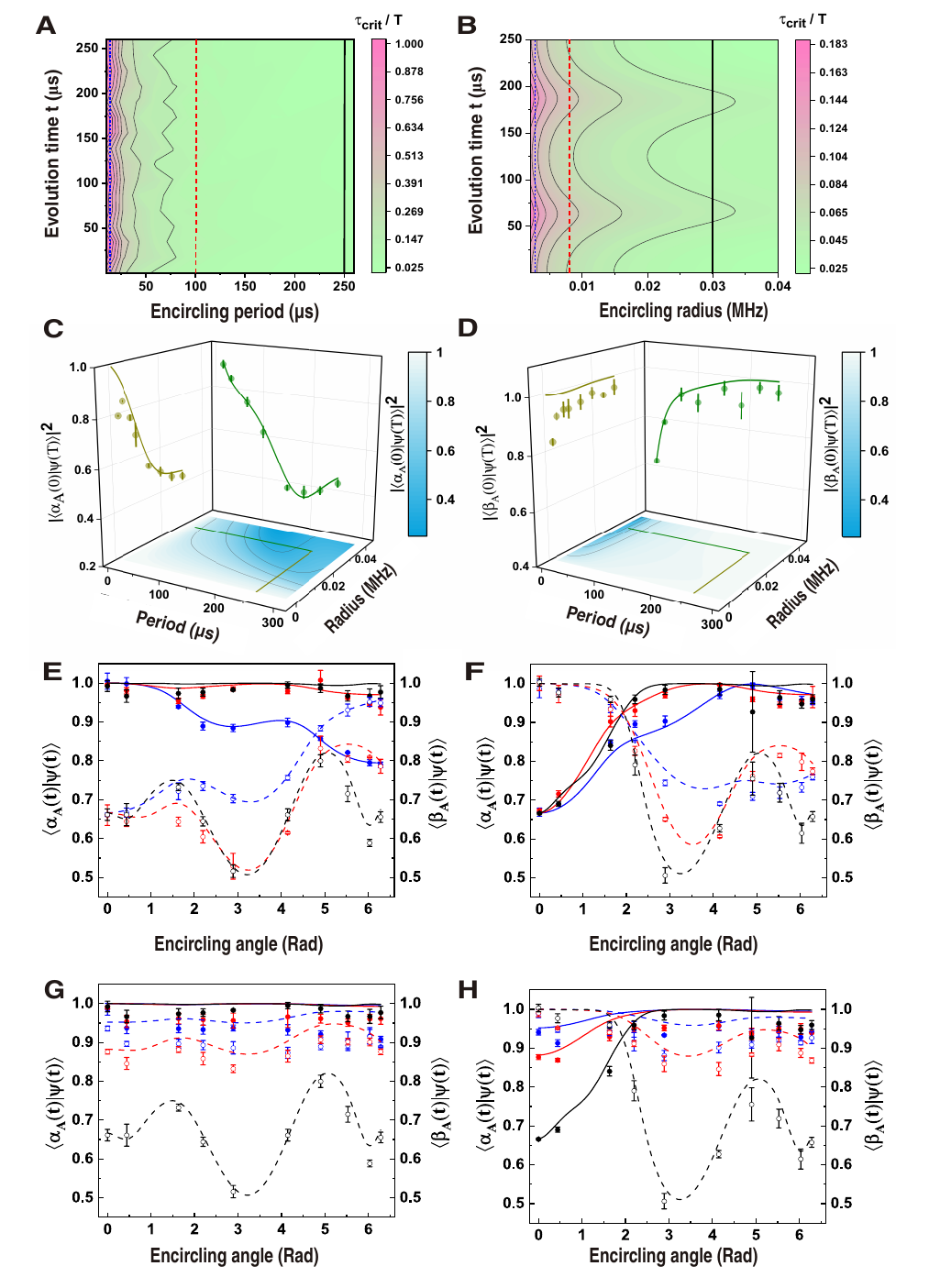}
	\caption{Investigation of the nonadiabatic effects with clockwise encircling the EP starting from the initial eigenstate $|\alpha_A(0)\rangle$ (trajectory 1). (\textbf{A-B}) The dependence of the ratio $\tau_{\textrm{crit}}/T$ on the encircling period and radius. (\textbf{C-D}) The change in the fidelity of state transfer with varying periods and radii. The contour map and solid lines denote the simulation results, while the circles represent the experimental data. (\textbf{E-F}) The overlap $\langle\alpha_{A}(t)|\psi(t)\rangle$ (solid line) and $\langle\beta_{A}(t)|\psi(t)\rangle$ (dash line) with varying period for the initial state (E) $|\alpha_A(0)\rangle$ and (F) $|\beta_A(0)\rangle$, respectively. The blue, red, and black circles correspond to $T=16.67~\mu s$, $100~\mu s$, and $250~\mu s$, respectively, which match the vertical lines in (A). (\textbf{G-H}) The overlap with varying radii for the initial state (E) $|\alpha_A(0)\rangle$ and (F) $|\beta_A(0)\rangle$, respectively. The blue, red, and black squares correspond to $r =$0.003 MHz, 0.008 MHz, and 0.03 MHz, respectively, which correspond to the vertical lines in (B).}	
	\label{ptrt}
\end{figure*}

We have developed a quantitative approach to examine the dependence of chirality (nonreciprocity) on the adiabaticity of the encircling process. In this method, the chiral (nonreciprocal) behavior is characterized by the the fidelity of state transfer during one encircling of EP, defined as the squared overlap between the final state $\psi(T)$ and the initial state. The adiabaticity of the encircling is quantified by $\tau_{\textrm{crit}}$= max$(\frac{1}{\lambda_{1}(t)-\lambda_{2}(t)})$. In the experiments, the fidelities of state transfer $|\alpha_A(0)\rangle \xrightarrow{\text{CW}} |\beta_A(0)\rangle$  (trajectory 1), i.e., $|\langle \alpha_A(0)|\psi(T)\rangle|^2$ and $|\langle \beta_A(0)|\psi(T)\rangle|^2$, are studied with various encircling periods and radii, as shown in Fig.~\ref{ptrt}. These results demonstrate the dependence of chirality (nonreciprocity) on the adiabaticity of the encircling process.

The results for various periods are shown in Fig.~\ref{ptrt}A. $\tau_{\textrm{crit}}=11.8~\mu s$ for all selected periods. For an encircling period $T=250~\mu$s $\gg \tau_{\textrm{crit}}$ that satisfies the adiabatic criteria,
the fidelity of state transfer $\langle\beta_A(0)|\psi(T)\rangle^2$ is nearly unity, representing almost perfect state swaps of $|\alpha_A(0)\rangle$ and $|\beta_A(0)\rangle$. When the encircling period reduces, the deviations from the ideal fidelities get larger and larger, as shown in shown in the olive curve of the side plane in Fig.~\ref{ptrt}, C-D. For example, for $T=100~\mu$s, $\langle\beta_A(0)|\psi(T)\rangle$ becomes smaller than the unity, indicating that the final state fails to fully reach $|\beta_A(0)\rangle$. This decay of state overlap is more obvious for $T=16.67~\mu $s. Such deviation from $|\beta_A(0)\rangle$ lead to the breakdown of the chiral (nonreciprocal) state transfer.

The results for various radii are depicted in the dark yellow curve on the side plane in Fig.~\ref{ptrt}, C-D, showing chiral (nonreciprocal) behavior primarily observed at larger radii. For $r=$0.03 MHz, the fidelity of $\langle\beta_A(0)|\psi(T)\rangle^2$ approaches unity. However, as the radius decreases, $\tau_{crit}/T < 0.1$, resulting in a significant deviation from the predictions of the adiabatic theorem. The state transfer for the $\mathcal{APT}$-symmetric Hamiltonian, with varying periods and radii, can be found in Supplemental Material S2.

Here, we observe DNATs during the adiabatic evolution of system parameters, which are distinctly different from the conventional speed-dependent nonadiabatic transitions (SNATs) where the evolution does not satisfy the adiabatic theorem. In Fig.~\ref{ptrt}E, as the period reduces from $T=250~\mu$s to $T=16.67~\mu$s, we observe a crossing of $\langle\alpha_{A}(t)|\psi(t)\rangle$ and $\langle\beta_{A}(t)|\psi(t)\rangle$ (for $T=16.67~\mu$s blue curves), indicating the occurrence of a SNAT.
Additionally, reducing the encircling period deviates the path from the prediction dictated by the adiabatic theorem, as demonstrated by the different crossings in Fig.~\ref{ptrt}F. The new crossing of $\langle\alpha_{A}(t)|\psi(t)\rangle$ and $\langle\beta_{A}(t)|\psi(t)\rangle$ does not appear due to the dominant DNAT. Notably, decreasing the encircling radius does not generate SNATs but only decreases the fidelity of the DNAT, as shown in Fig.~\ref{ptrt}, G-H.

\subsection*{Robustness of the state transfer}
\begin{figure*}
	\centering
	\includegraphics[width=0.7\textwidth]{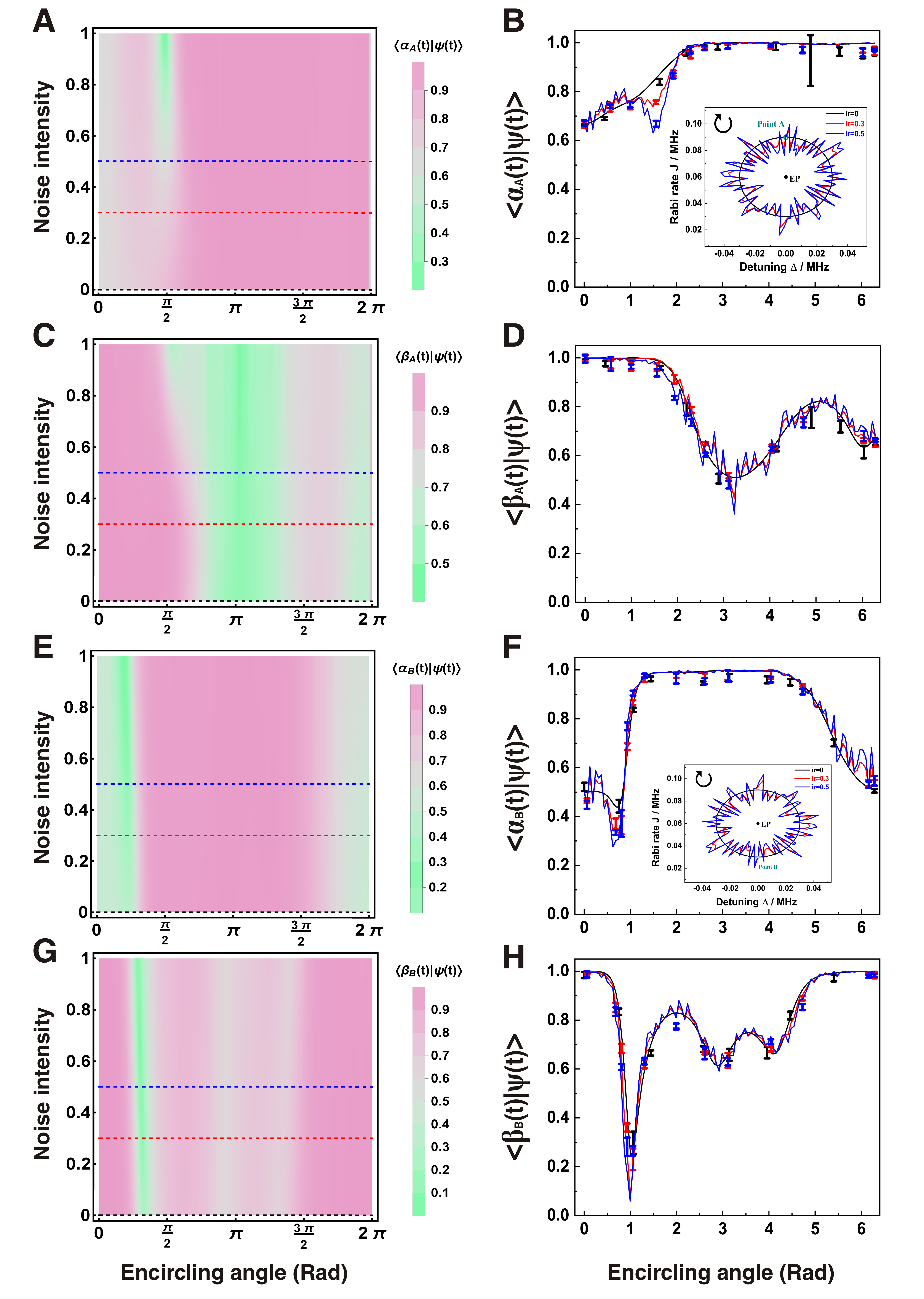}
	\caption{Investigation of the robustness of dynamically encircling an EP.
		(\textbf{A, C, E and G}) The color map shows the overlap between the instantaneous eigenstate and the evolutionary state under varying noise intensity. (\textbf{B, D, F, and H}). The experimental results corresponding to the three dashed lines in (A, C, E, and G) demonstrate robustness against noise, as all data match the predicted values (solid lines). The insets in (B) and (F) display clockwise encircling trajectories in the two-dimensional parameter space $\{\Delta, J\}$ with random noise intensities $ir =$ 0 (black), 0.3 (red), and 0.5 (blue), respectively. In Fig.~A-D, the encircling starts from point A in the $\mathcal{PTS}$ regime, while in Fig.~E-H, it starts from point B in the $\mathcal{PTB}$ regime. }
	\label{PTN}
\end{figure*}
We experimentally verify the topological robustness of state transfers using $\mathcal{PT}$ and $\mathcal{APT}$-symmetric Hamiltonians against the physical noises in a dissipative trapped-ion qubit. For the $\mathcal{PT}$ Hamiltonian, we introduce noises into the detuning and the coupling rate along the encircling path
\begin{equation}
	\label{index3}
	\begin{cases}
		\Delta(t) = r (1+\kappa(t)) Sin[\omega t +\theta_0] \\
		J(t) = \gamma + r (1+\kappa(t)) Cos[\omega t +\theta_0],
	\end{cases}
\end{equation}
where the encircling radius $r = 0.03$ MHz, the dissipation rate $\gamma = 0.06 $ MHz, and $\kappa(t)$ represents the noise in the encircling process. $\kappa(t)\in[-ir, ir]$ is generated by pseudorandom real number, where $ir$ is defined as the intensity of the noise.

Figure.~\ref{PTN} illustrates the final state outcomes of the encircling process, showing that it remains unaffected by random noise, regardless of the noise intensity. Figs.~\ref{PTN}, A to D (E to H) display the experimental results for the $\mathcal{PT}$-symmetric (broken) regime. Remarkably, we find that chiral (nonreciprocal) quantum state transfers exhibit robustness against such noise, provided the adiabatic encircling condition is satisfied. Even in the adiabatic limit of evolution, this robustness persists near the EP because the external noise fails to eliminate the degeneracy at the EP. We have also experimentally verified the topological robustness of the quantum state transfer with the $\mathcal{APT}$-symmetric Hamiltonian, as detailed in Supplementary Material S2. In both cases, the experimental data match pretty well with the theoretically predicted values, even under varying noise intensities. This validation confirms the robustness of topological state transfer with $\mathcal{APT}$-symmetric Hamiltonians.

\section*{Discussion}
In this study, we have observed chiral (nonreciprocal) quantum dynamics in a trapped-ion qubit, wherein the EPs of both $\mathcal{PT}$-symmetric and $\mathcal{APT}$-symmetric Hamiltonians are dynamically encircled. These dynamics are rooted in the unique topological structure of  intersecting Riemann sheets around the EP. During adiabatic state transfers, this structure leads to the swapping of two eigenstates after one cycle and the acquisition of a geometric phase of $\pi$ after two cycles~\cite{dembowski2001experimental,ding2016emergence}. The topological invariant of the adiabatic process is characterized by this  geometric phase, providing the protection against perturbations. For nonadiabatic processes, it is usually believed that the nonadiabaticity would cause the evolution to deviate from the eigenstates of the instantaneous Hamiltonian, and consequently undermine topological protection from the Riemann surface.
However, previous experiments have confirmed the robust state transfer not only for adiabatic processes, but also existing in  ones involving dissipation-induced nonadiabatic transitions (DNATs)~\cite{gao2015observation,zhou2018observation,tang2020exceptional,yang2023realization}. This raises one critical question: is there a topological invariant to elucidate the  robust state transfers in nonadiabatic processes , which could unveil the topological structure of  DNATs. Addressing this question is crucial for fully understanding chiral and nonreciprocal state transfers in non-Hermitian systems. To date, this question remains unexplored.

Based on the above results, we provide a scheme to address this question, in which a new topological invariant that we dub as ``dynamic vorticity" is used to characterize
topological protections in the processes involving DNATs.  Previously, the vorticity $\mathcal{V}=-\frac{1}{2\pi}\oint_{C}\nabla_{\theta}Arg[E_{+}(\theta)-E_{-}(\theta)]\cdot d\theta$, also known as the spectral or eigenvalue winding number\cite{su2021direct,ding2022non}, has been applied to explain the topological protection of the adiabatic state transfers ~\cite{shen2018topological,kawabata2018anomalous}. $C$ is a closed loop in the complex energy plane, $\theta$ is the encircling angle, $E_{\pm}$ are the eigenvalue of the Hamiltonian. Remarkably, we find that this concept can be extended to the process involving DNATs, if we adopt the parallel transported eigenbasis~\cite{berry1984quantal}. This transformation into the parallel transported eigenbasis gives an effective time-dependent Hamiltonian
\begin{equation}
	\widetilde{ H}_{eff}(t)=  \begin{pmatrix}
		-i\gamma -\lambda(t) & f(t)\\
		-f(t) &  -i\gamma + \lambda(t)
	\end{pmatrix},
	\label{newHeff}
\end{equation}
where $f(t)$ is the term corresponding to the nonadiabatic transition (Details in Materials and methods). By extending the definition of the vorticity to the parallel transported eigenbasis, the dynamic vorticity is given by
\begin{equation}
	\mathcal{V_D}=-\frac{1}{2\pi}\oint_{C}\nabla_{\theta}\textrm{Arg}[\widetilde{E}_{+}(\theta)-\widetilde{E}_{-}(\theta)]\cdot d\theta,
\end{equation}
where $\widetilde{E}_{\pm}$ is the eigenvalue of $\widetilde{ H}_{eff}(t)$. The dynamic vorticity associates with the energy dispersion of $\widetilde{ H}_{eff}(t)$, and  inherits the topology of Riemann surfaces. The essence of its validity lies in the fact that nonadiabatic transitions will not cause the deviations of the instantaneous state from the Riemann surfaces in the parallel transported basis. Consequently, the system's evolution consistently remains in the eigenstate of $\widetilde{ H}_{eff}(t)$. Therefore, it can be treated as a universal topological invariant for dissipation-induced nonadiabatic state transfers near the EP.

In our experiments shown in Fig.~\ref{PTN}, we find that $\mathcal{V_D}=\pm 1/2$, i.e., $-1/2$ for  trajectory 1 and 1/2 for the trajectory 2, confirming the chiral (nonreciprocal) state transfer is topologically protected. It is further confirmed by the state transfers investigated in Refs~\cite{zhang2018dynamically,nasari2022observation}, where $\mathcal{V_D}=0$ characterizes the topological property of the processes that do not encircle an EP.  From theoretical calculations,  we conclude that $\mathcal{V_D}=\pm n/2$, where $n$ represents the number of EPs being encircled and the sign $\pm$ depends on the orientation of the encircling curve.

Another advantage of using the parallel transported eigenbasis is that the effective Hamiltonian in this basis can be used to determine the occurrence of chiral behavior. If the initial effective Hamiltonian satisfies $\{\widetilde{H}_{eff}(0),\mathcal{CPT} \}=0$, where we define the chiral operator $\mathcal{S}=\mathcal{CPT}$, our experimental results indicate a pair of encircling dynamics
have the chiral symmetry (i.e., trajectories 1 and 2 in Fig.~\ref{3D-PTS} have different final states). Here, parity operator $\mathcal{P}=\sigma_{x}$, $\mathcal{C}=\sigma_{z}$, and time-reversal operator $\mathcal{T}$ represents the conjugate operator. Conversely, if $\{\widetilde{H}_{eff}(0),\mathcal{CPT} \}\neq 0$, the dynamics breaks the chiral symmetry (i.e., trajectories 5 and 6 in Fig.~\ref{3D-PTS} have the same final states). Note that $\mathcal{S} = \mathcal{CPT}$ instead of $\mathcal{S} = \mathcal{CT}$ in the Hermitian case~\cite{chiu2016classification, qi2018defect} due to non-Hermiticity~\cite{bender2002complex}. We also find that the nonreciprocal behavior depends on the initial density matrix $\rho(0)$. The positive (negative) value of $\textrm{Tr}(\sigma_z\rho(0)) $ corresponds to the preservation (breakdown) of TRS for a pair of the forward-time and backforward-time encircling processes. When $\textrm{Tr}(\sigma_z\rho(0)) =0$, if $d[\textrm{Tr}(\sigma_z\rho(t))]/dt|_{t=0}> 0$, the processes preserve the TRS (i.e., trajectories 1 and 4). Otherwise, they exhibits nonreciprocal behavior. Therefore, we conclude that the chiral and nonreciprocal dynamics are dictated by the symmetry of the effective Hamiltonian $\widetilde{ H}_{eff}(0)$ and the initial state.

It is worth mentioning that the dynamic vorticity is limited to the dissipation-induced nonadiabatic processes and is not suitable for speed-induced nonadiabatic processes, where the evolution period is shorter than the critical time stipulated by the adiabatic theorem. Our results in Fig.~\ref{ptrt} show that the occurrence of SNATs disrupts the chiral and nonreciprocal dynamics due to the nonadiabaticity of the Landau-Zener tunneling from an eigenstate to another at the avoided-crossing.

\section*{Materials and methods}
\subsection*{Encircling method}
The dynamically encircling of the EP is realized by using the time-dependent detuning $\Delta(t)= r \sin[\theta (t) +\theta_0] $ (in MHz) and the time-dependent coupling rate $J(t)= J_0+ r \cos[\theta(t) +\theta_0]$ (in MHz), while keeping the dissipation rate fixed~\cite{liu2021dynamically}. Here, $r$ is the encircling radius, $\theta(t)=\omega t$ is the encircling angle, and $\omega$ is the angular speed of encircling whose sign determines the encircling direction (``+" for clockwise encircling and ``-" for counterclockwise encircling). The parameter $\theta_0=0$ ( $\theta_0=\pi$) indicates that the starting point lies at the $\mathcal{PTS}$ ($\mathcal{PTB}$) regime.

The evolutionary state at each time step, i.e. $|\psi(t)\rangle = C_{1}(t)|\alpha(t)\rangle + C_{2}(t)|\beta(t)\rangle$ is the coherent superposition of the eigenstates $|\alpha(t)\rangle$ and $|\beta(t)\rangle$. The coefficients $C_{1}(t)$ and $C_{2}(t)$ are probability amplitudes of the $|\alpha(t)\rangle$ and $|\beta(t)\rangle$, respectively. The trajectory of the encircling, i.e. $(|C_{1}(t)|^2 \lambda_{1} +|C_{2}(t)|^2 \lambda_{2})/(|C_{1}(t)|^2+|C_{2}(t)|^2)$ is derived, where the sudden transitions between $|\alpha(t)\rangle$ and $|\beta(t)\rangle$ can be analyzed through the adiabatic multipliers~\cite{wang2018non,berry2011optical}. With the derived trajectory, the state evolution is projected onto the complex Riemann sheets of the eigenvalues, shown as the black line in Fig.~1B of the main text.

The presence of spin-dependent loss causes the qubit population to exponentially decay in a sinusoidal manner for small $\gamma/J$ ratios~\cite{lu2024realizing,bian2023quantum,ding2021experimental}, leading to an extremely low state population that is challenging to detect experimentally. In response to this challenge, we employ a piecewise strategy. Encircling starts at $t = 0$ and ends at $T = \frac{15}{\gamma}$. Throughout the encircling process, the dissipation rate remains constant at $\gamma=0.06$ MHz, the angular speed is maintained at $\omega = \pm \frac{2\pi}{T}$, and the radius is fixed at $r=0.03$ MHz (unless specified otherwise in subsequent experiments).
%\subsection*{Piecewise strategy}
\begin{figure}[h]
	\includegraphics[width=1.0\linewidth]{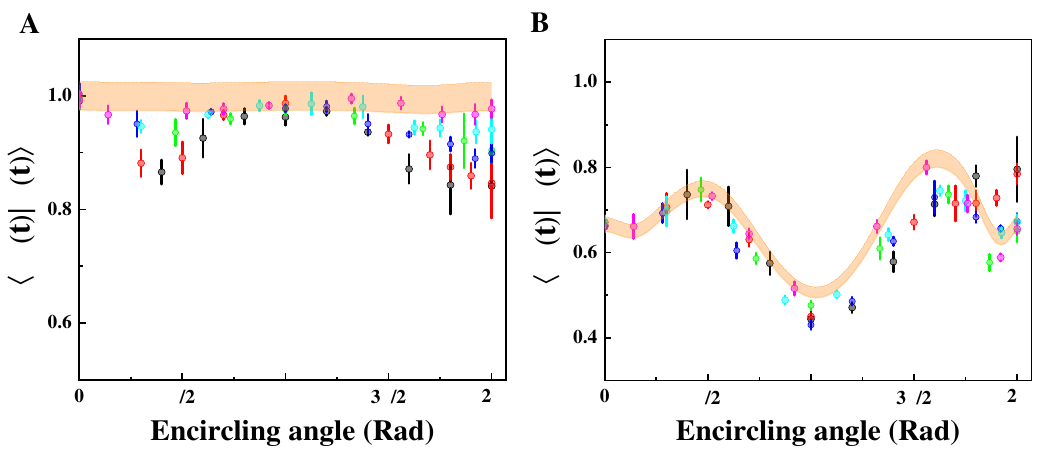}
	\caption{Experimental verification of the piecewise strategy with the initial eigenstate $|\alpha_A(0)\rangle$ in the $\mathcal{PT}$-symmetric regime. (A) The overlap between the evolutionary state $|\psi(t)\rangle$ and instantaneous eigenstate $|\alpha_A(t)\rangle$. (B) The overlap between the evolutionary state $|\psi(t)\rangle$ and instantaneous eigenstate $|\beta_A(t)\rangle$. The experimental results with $N=$10, 20, 30, 50, 80 and 100 are represented by black, red, green, blue, cyan, and magenta squares, respectively. The yellow shaded region represents the numerical calculation with $5\%$ uncertainty in the density matrix of the measured state $|\psi(t)\rangle$, accounting for fluctuations in the dissipation strength and the coupling strength.}
	\label{verification}
\end{figure}

The entire encircling path, with a period of $T=2\pi/\omega$, is divided into $N$ segments. In each of the $N$ segments ($1\le n\le N$), the qubit state is prepared to evolve for $t_n=(n-1)T/N$ time according to theoretical prediction. Then, it continues to evolve for $t=T/N$ under the Hamiltonian $H(t_n)$. With this scheme, we map out the whole encircling process by quantum state tomography of $|\psi(t)\rangle$, which is carried out at the end point of each segment. The overlap (i.e., the inner product) of the measured evolutionary state $|\psi(t)\rangle$ and the instantaneous eigenstates $|\alpha_{A}(t)\rangle$ (or $|\beta_{A}(t)\rangle$) of the time-varying Hamiltonian $H_{eff}(t)$ is used to evaluate whether the measured state $|\psi(t)\rangle$ matches with the theoretical calculation.

We adjust the size of $N$ to verify the effectiveness of the piecewise strategy as shown in Fig.~\ref{verification}, where $J=0.06$ MHz and $\omega=2\pi\times$ 4 rad ms$^{-1}$, respectively. The encircling starts from $|\alpha_A(0)\rangle$, and $N$ varies with 10 (black squares), 20 (red squares), 30 (green squares), 50 (blue squares), 80 (cyan squares) and 100 (magenta squares).  The numerical calculation has a $5\%$ uncertainty (orange error band) in the density matrix of the measured evolutionary state $|\psi(t)\rangle$, accounting for the fluctuations of the dissipation strength and the coupling strength. The nonzero overlap between $|\beta_A(0)\rangle$ and $|\psi(0)\rangle$ is due to the nonorthogonality of the two eigenstates of $H_{eff}(0)$. The experimental results for $N=100$ agree well with the numerical simulation, while the others do not, proving the validness of the proposed scheme for $N\ge 100$. Therefore, we choose $N=100$ in the following experiments to investigate the dynamics of the encircling.

\subsection*{Mapping nonadiabatic transition amplitudes}
\begin{figure*}
	\centering
	\includegraphics[width=0.7\linewidth]{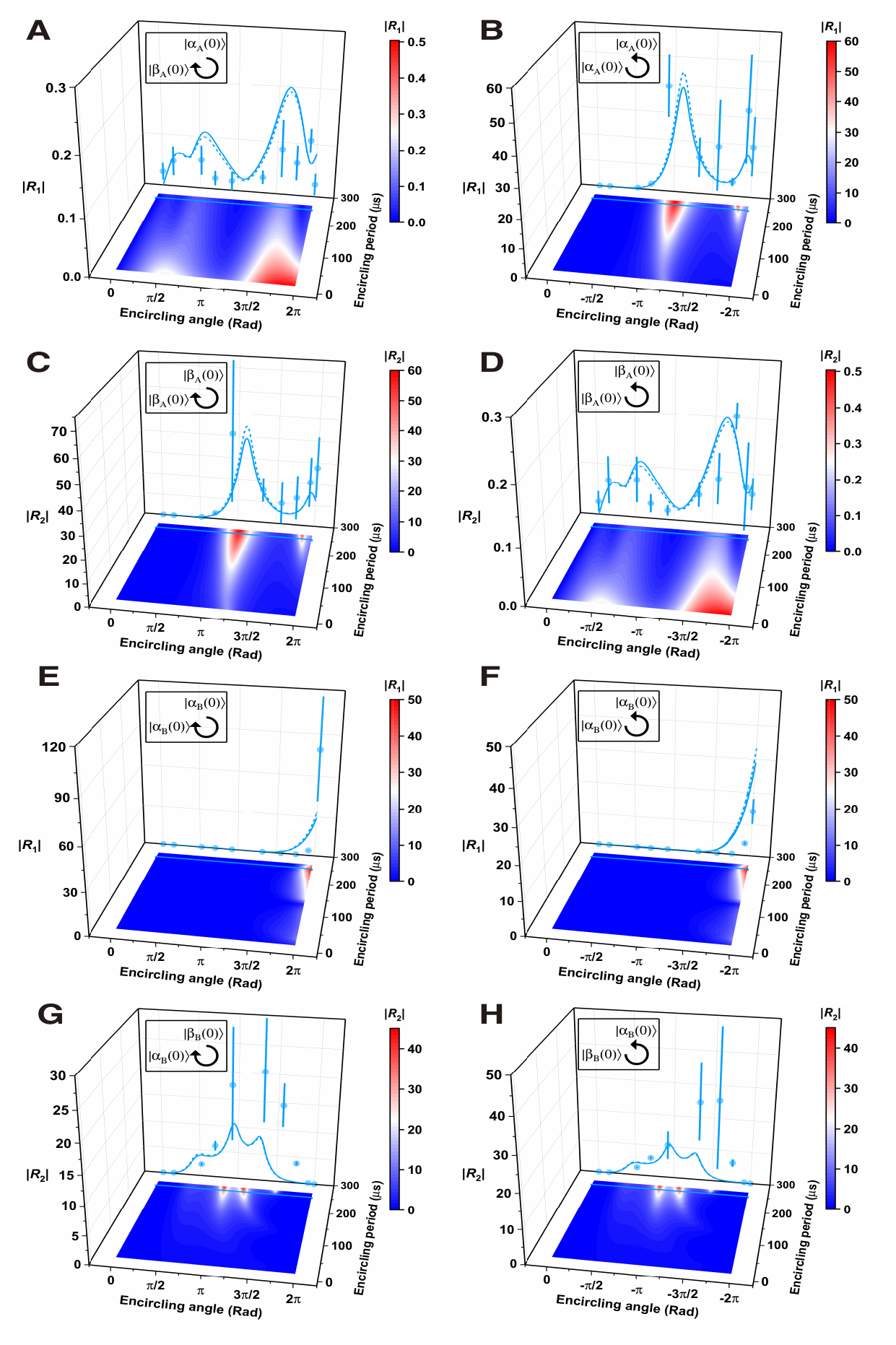}
	\caption{The time-dependent nonadiabatic transition amplitude when the $\mathcal{PT}$ Hamiltonian encircles the EP. (A-D) Encircling starting from the $\mathcal{PT}$-symmetric regime:
		(A) Clockwise with the initial state $|\alpha_A(0)\rangle$,
		(B) Counterclockwise with the initial state $|\alpha_A(0)\rangle$,
		(C) Clockwise with the initial state $|\beta_A(0)\rangle$,
		(D) Counterclockwise with the initial state $|\beta_A(0)\rangle$.
		(E-H) Encircling starting from the $\mathcal{PT}$-broken regime:
		(E) Clockwise with the initial state $|\alpha_B(0)\rangle$,
		(F) Counterclockwise with the initial state $|\alpha_B(0)\rangle$,
		(G) Clockwise with the initial state $|\beta_B(0)\rangle$,
		(H) Counterclockwise with the initial state $|\beta_B(0)\rangle$.
		The color maps for the time-dependent nonadiabatic transition amplitude when changing the encirlcing period. The cyan circles with error bars, solid and dashed lines on the  encircling angle-$|R_i|$ plane correspond to the experimental data, theoretical calculation (Eq.~\ref{rde}) and numerical simulation from $C_1(t)$ and $C_2(t)$, where the encircling period $T=250~\mu s$.}
	\label{ptr}
\end{figure*}
The nonadiabatic transitions that occur during the encircling of the EP can be attributed to the Stokes phenomenon of asymptotics~\cite{olde1995stokes,berry2011slow} or stability loss delay~\cite{milburn2015general,feilhauer2020encircling}. Here, we analyze DNATs observed in our experiment using the method of parallel transported basis~\cite{milburn2015general}. The state evolution for a passive $\mathcal{PT}$ Hamiltonian is determined by ($\hbar$=1)
\begin{equation}
	i\frac{\partial}{\partial t} \left | \psi\left ( t \right )   \right \rangle=H_{eff}\left | \psi\left ( t \right )   \right \rangle ,
	\label{S-ion-loss}
\end{equation}
where $\left | \psi(t)  \right \rangle=C_1(t)\left | \alpha (t)  \right \rangle +C_2(t)\left | \beta (t)  \right \rangle$ and $H_{eff}= \begin{pmatrix}
	\Delta/2  & J \\
	J & -\Delta/2-2i\gamma
\end{pmatrix}$.

The eigenvalues of $H_{eff}$ are
$\widetilde{\lambda}_{\pm} = -i\gamma \pm \sqrt{(\Delta/2 + i \gamma)^2+J^2} = -i\gamma \pm \lambda$,
where $\lambda$ is the eigenvalue of the $\mathcal{PT}$-symmetric Hamiltonian $H_{\mathcal{PT}}$ with detuning. The normalized eigenstates of $H_{eff}$ are
\begin{equation}
	\begin{split}
		\chi_1 = \frac{1}{\sqrt{(\lambda-(\Delta/2 + i \gamma))^2+J^2}}\begin{pmatrix}
			-\lambda+(\Delta/2 + i \gamma) \\
			J
		\end{pmatrix}\\
		\chi_2 = \frac{-1}{\sqrt{(\lambda+(\Delta/2 + i \gamma))^2+J^2}}\begin{pmatrix}
			\lambda+(\Delta/2 + i \gamma) \\
			J
		\end{pmatrix},
	\end{split}
\end{equation}
which can be expressed as the parallel transported eigenbasis
\begin{equation}
	\left | \alpha  \right \rangle = \chi_1 = \begin{pmatrix}
		cos(\theta/2) \\
		sin(\theta/2)
	\end{pmatrix}, \\
	\left | \beta  \right \rangle =\chi_2 = \begin{pmatrix}
		-sin(\theta/2) \\
		cos(\theta/2)
	\end{pmatrix},
	\label{transported-eigenbasis}
\end{equation}
where
\begin{equation}
	\begin{split}
		\sin(\theta/2) = \sqrt{\frac{\lambda+(\Delta/2 + i \gamma)}{2\lambda}}\\
		\cos(\theta/2) = -\sqrt{\frac{\lambda-(\Delta/2 + i \gamma)}{2\lambda}}\\
		\tan\theta = J/(\Delta/2 + i \gamma).
	\end{split}
\end{equation}
Then, we use $\left | \alpha  \right \rangle=T \begin{pmatrix}
	1 \\0
\end{pmatrix}$, 
$\left | \beta  \right \rangle=T \begin{pmatrix}
	0 \\1
\end{pmatrix}$ to represent the parallel transported eigenbasis of Eq.~\ref{transported-eigenbasis}.
The rotation matrix $T = \begin{pmatrix}
	\cos (\theta/2) & -\sin (\theta/2) \\
	\sin (\theta/2) & \cos (\theta/2)
\end{pmatrix}$,
where $TT^\top = T^\top T = 1$. 

Now $|\psi(t)\rangle = T|\psi'(t)\rangle$, where 	$|\psi'(t)\rangle = C_1(t)\begin{pmatrix}
	1\\0
\end{pmatrix} + C_2(t)\begin{pmatrix}
	0\\1
\end{pmatrix} = \begin{pmatrix}
	C_1(t)\\C_2(t)
\end{pmatrix}$.
Substituting the $|\psi(t)\rangle = T|\psi'(t)\rangle$ into  Eq.~\ref{S-ion-loss}, we obtain
\begin{equation}
	i \frac{\partial}{\partial t}(T|\psi'(t)\rangle) = H_{eff}(T|\psi'(t)\rangle).
	\label{sch}
\end{equation}

We consider the evolution $U'(t)$ defined by $|\psi'(t)\rangle = U'(t) |\psi'(0)\rangle$. Eq.~\ref{sch} can be rewritten as $\frac{\partial}{\partial t}(TU'(t))= -i H_{eff} (T U'(t))$, which is simplified as
\begin{equation}
	\dot{T} U'(t) + T \dot{U'}(t)= -i H_{eff} (T U'(t)).
	\label{U'-ion-loss}
\end{equation}
Multiplying $T^\top$ to the left of both sides in Eq.~\ref{U'-ion-loss}, we obtain
\begin{equation}
	T^\top\dot{T} U'(t) + \dot{U'}(t)= -i (T^\top H_{eff} T) U'(t),
	\label{Ex-ion-loss}
\end{equation}
where
\begin{equation}
	\dot{T}  = \frac{\lambda(\dot{\Delta}/2+i\dot{\gamma})-\dot{\lambda}(\Delta/2+i\gamma)}{(2\lambda)^2}\begin{pmatrix}
		-\frac{1}{\cos (\theta/2)} & -\frac{1}{\sin (\theta/2)} \\
		\frac{1}{\sin (\theta/2)} & -\frac{1}{\cos (\theta/2)}
	\end{pmatrix}.
\end{equation}
Then,
\begin{equation}
	\begin{split}
		T^\top\dot{T} = \frac{\lambda(\dot{\Delta}/2+i\dot{\gamma})-\dot{\lambda}(\Delta/2 + i \gamma)}{2\lambda J}\begin{pmatrix}
			0 & 1\\
			-1& 0
		\end{pmatrix}\\
		T^\top H_{eff} T = \begin{pmatrix}
			-i\gamma - \lambda & 0\\
			0&  -i\gamma + \lambda
		\end{pmatrix}.
	\end{split}
	\label{TTT}
\end{equation}
Substituting Eq.~\ref{TTT} into Eq.~\ref{Ex-ion-loss}, we obtain
\begin{equation}
	\dot{U'}(t)= -i (-i T^\top\dot{T}+T^\top H_{eff}T)U'(t)=-i\widetilde{ H}_{eff} U'(t),
\end{equation}
where
\begin{equation}
	\begin{aligned}
		\widetilde{ H}_{eff}&= -iT^\top\dot{T}+T^\top H_{eff}T = \begin{pmatrix}
			-i\gamma -\lambda & f\\
			-f &  -i\gamma + \lambda
		\end{pmatrix}, \\
		f &= \frac{J(\dot{\Delta}/2+i\dot{\gamma})-\dot{J}(\Delta/2 + i \gamma)}{2i\lambda^2}.
	\end{aligned}
	\label{sf}
\end{equation}

Expressing $U'(t)$ in matrix form
$U'(t) = \begin{pmatrix}
	U'_{\alpha\alpha}(t) & U'_{\alpha\beta}(t)\\
	U'_{\beta\alpha}(t) & U'_{\beta\beta}(t)\\
\end{pmatrix}$. Its differential form can be described as
\begin{equation}
	\begin{split}
		\dot{U'}_{\alpha\alpha}(t)= -i((-i\gamma-\lambda) U'_{\alpha\alpha}(t) + f U'_{\beta\alpha}(t))\\
		\dot{U'}_{\alpha\beta}(t) = -i((-i\gamma-\lambda) U'_{\alpha\beta}(t) + f U'_{\beta\beta}(t))\\
		\dot{U'}_{\beta\alpha}(t) = -i((-i\gamma+\lambda) U'_{\beta\alpha}(t) - f U'_{\alpha\alpha}(t))\\
		\dot{U'}_{\beta\beta}(t) = -i((-i\gamma+\lambda) U'_{\beta\beta}(t) - f U'_{\alpha\beta}(t))
	\end{split}
	\label{dU}
\end{equation}

If the initial state is $|\alpha(0)\rangle$, meaning $C_1(0) = 1, C_2(0) = 0$, then the final state $
|\psi'(t)\rangle = U'(t)\begin{pmatrix}
	1\\0
\end{pmatrix} = \begin{pmatrix}
	U'_{\alpha\alpha}(t)\\U'_{\beta\alpha}(t)
\end{pmatrix}=\begin{pmatrix}
	C_1(t)\\C_2(t)\end{pmatrix}	$.
This implies that $U'_{\alpha\alpha}(t)$ and $U'_{\beta\alpha}(t)$ serve as the amplitudes of the corresponding eigenstates. Therefore,  the relative nonadiabatic transition amplitudes can be defined as $R_1(t) = \frac{U'_{\beta\alpha}(t)}{U'_{\alpha\alpha}(t)}$. Initially set at $R_1(0)=0$, if the evolution remains adiabatic, $R_1(t)<1$. However, in the case of a nonadiabatic transition, where two eigenstates exchange position, $R_1(t)>1$. The same applies to the initial state $|\beta\rangle$, where $C_1(0) = 0$ and $C_2(0) = 1$. In this case, $R_2(t) = \frac{U'_{\alpha\beta}(t)}{U'_{\beta\beta}(t)}$ is employed to indicate the occurrence of nonadiabatic transitions. Utilizing Eq.~\ref{dU}, we derive the differential equations governing the relative nonadiabatic transition amplitudes.
\begin{equation}
	\begin{split}
		\dot{R_1}(t) =  2i \lambda R_1(t) - i f (1+R^2_1(t))\\
		\dot{R_2}(t) = - 2i \lambda R_2(t)  + i f (1+R^2_2(t) ).
	\end{split}
	\label{rde}
\end{equation}
We can also calculate $R_1(t)$ ($R_2(t)$) by measuring $C_1(t)$ and $C_2(t)$, giving $R_1(t)=\frac{C_2(t)}{C_1(t)}$ ($R_2(t)=\frac{C_1(t)}{C_2(t)})$. For clockwise encircling from $|\alpha_A(0)\rangle$ in Fig.~\ref{ptr}(A),  $R_1(t) = \frac{C_2(t)}{C_1(t)}<1$ in the whole period, indicating $|C_1(t)|$ and $|C_2(t)|$ do not cross, so no DNAT occurs. Similarly, in Fig.~\ref{ptr}(B), $R_1(t) = \frac{C_2(t)}{C_1(t)}$ evolves from $R_1(t)<1$ to $R_1(t)>1$, indicating $C_1(t)$ and $C_2(t)$ cross, hence a DNAT occurs. Both the theoretical and experimental results for $\mathcal{PTS}$ and $\mathcal{PTB}$ regimes are shown in Fig.~\ref{ptr}.

\section*{Acknowledgments}
\noindent \textbf{Acknowledgments:}~We thank Profs.~Shidong Liang and Wei Yi for valuable discussions.
\textbf{Funding:} This work is supported by the Key-Area Research and Development Program of Guang Dong Province under Grant No.2019B030330001, the National Natural Science Foundation of China under Grant No.11774436, No.11974434 and No.12074439, Natural Science Foundation of Guangdong Province under Grant 2020A1515011159, Science and Technology Program of Guangzhou, China 202102080380, the Shenzhen Science and Technology Program under Grant No.2021Szvup172 and JCYJ20220818102003006. Le Luo acknowledges the support from Guangdong Province Youth Talent Program under Grant No.2017GC010656. Ji Bian acknowledges the support from the China Postdoctoral Science Foundation under Grant No.2021M703768.
\textbf{Author contributions:}
L.L. conceived the original idea and supervised the overall research.
P.L, Y.L, Q.L, T.L, and L.L. designed the experiments. P.L, Y.L, H.W and F.Z contributed the spin-dependent dissipation technique. P.L, J.B, and Y.L conducted the experimental work. P.L and Y.L analyzed data. Q.L, P.L, T.L, Y.L, and J.B carried out theoretical modeling of encircling. Y.L, P.L, Q.L, T.L and L.L prepared the manuscript. All authors provided helpful discussion.
\textbf{Competing interests:} The authors declare no competing interests.
\textbf{Data and materials availability:}All data are available in the main text or the supplementary materials.

%\appendix
%\renewcommand\thefigure{\thesection.\arabic{figure}}
%\section{Experimental setup}
%\setcounter{figure}{0}

%%%%%%%%%% Merge with supplemental materials %%%%%%%%%%
%\pagebreak
\clearpage
\widetext

\begin{center}
\textbf{\large {Dynamical topology of chiral and nonreciprocal state transfers in a non-Hermitian quantum system}}
\end{center}
%%%%%%%%%% Merge with supplemental materials %%%%%%%%%%
%%%%%%%%%% Prefix a "S" to all equations, figures, tables and reset the counter %%%%%%%%%%
\setcounter{equation}{0}
\setcounter{figure}{0}
\setcounter{table}{0}
\setcounter{page}{1}
\setcounter{section}{0}
\makeatletter
\renewcommand{\theequation}{S\arabic{equation}}
\renewcommand{\thefigure}{S\arabic{figure}}
\renewcommand{\thetable}{S\arabic{table}}
\renewcommand{\bibnumfmt}[1]{[S#1]}
\renewcommand{\citenumfont}[1]{S#1}
\renewcommand{\thesection}{S\arabic{section}}

\section*{S1~Dynamically encircling of $\mathcal{PT}$ Hamiltonian}
\subsection*{S1.1 Time-varying nonadiabatic transition amplitude}
\begin{figure*}
	\centering
	\includegraphics[width=0.9\textwidth]{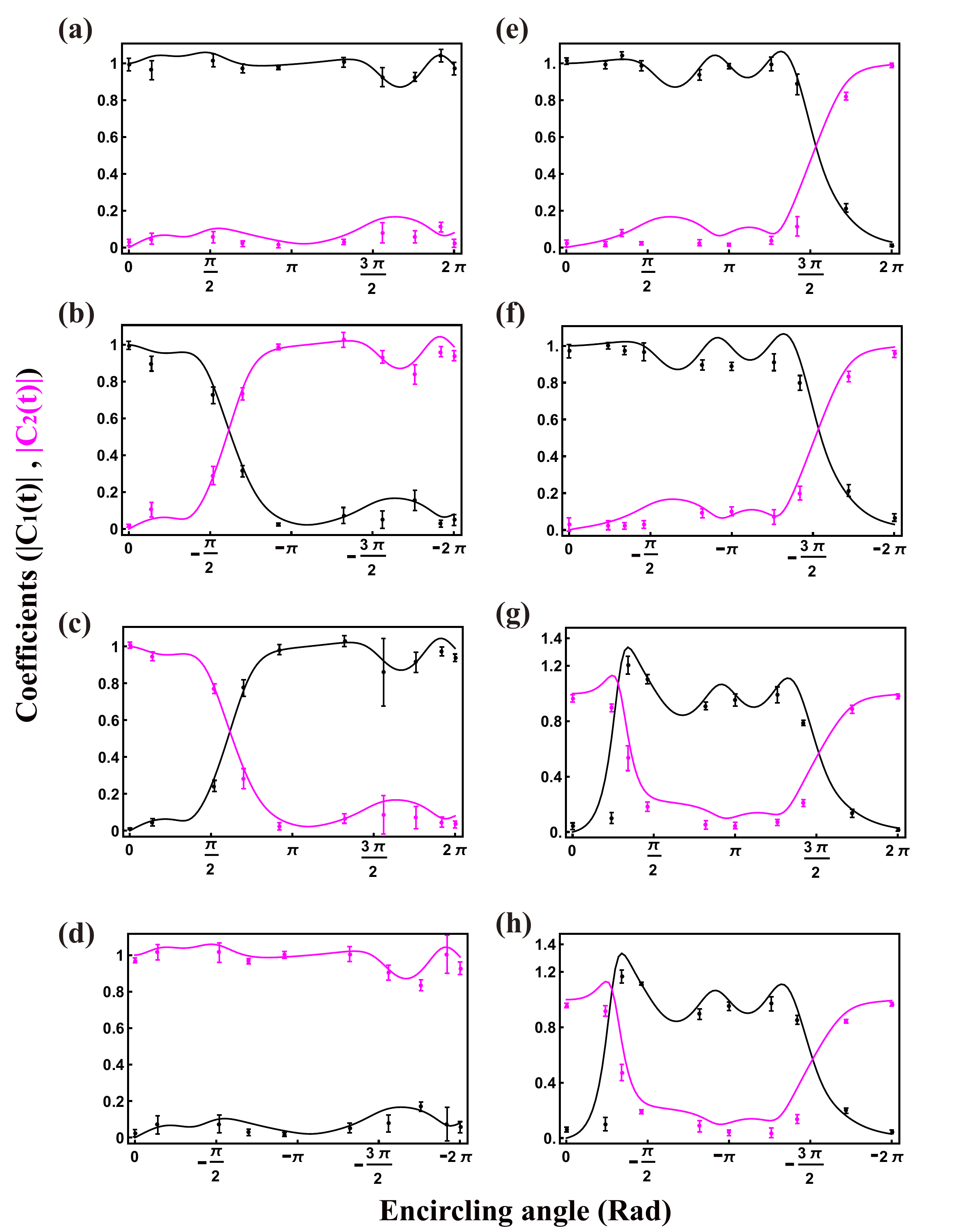}
	\caption{The amplitude coefficients of the instantaneous eigenstates when the $\mathcal{PT}$ Hamiltonian encircles the EP.
		(a-d) Encircling starting from the $\mathcal{PT}$-symmetric regime, with
		(a) Clockwise with the initial state $|\alpha_A(0)\rangle$,
		(b) Counterclockwise with the initial state $|\alpha_A(0)\rangle$,
		(c) Clockwise with the initial state $|\beta_A(0)\rangle$,
		(d) Counterclockwise with the initial state $|\beta_A(0)\rangle$.
		(e-h) Encircling starting from the $\mathcal{PT}$-broken regime, with
		(e) Clockwise with the initial state $|\alpha_B(0)\rangle$,
		(f) Counterclockwise with the initial state $|\alpha_B(0)\rangle$,
		(g) Clockwise with the initial state $|\beta_B(0)\rangle$,
		(h) Counterclockwise with the initial state $|\beta_B(0)\rangle$.}
	\label{ptc}
\end{figure*}

The nonadiabatic transition amplitudes with a $\mathcal{PT}$-symmetric Hamiltonian have been measured in our experiments, as depicted in Fig.~\ref{ptc}.  The time-dependent amplitude coefficients of $C_1(t)$ and $C_2(t)$ are determined by $|\psi(t)\rangle = C_1(t)|\alpha(t)\rangle + C_2(t)|\beta(t)\rangle$. The crossing of  $C_1(t)$ and $C_2(t)$ serves as a characterization of the nonadiabatic transition. According to stability loss delay, the time-varying relative nonadiabatic transition amplitude $R_1(t)$ and $R_2(t)$ can also be obtained by measuring $C_1(t)$ and $C_2(t)$, Specifically, these are given by  $R_1(t)=\frac{C_2(t)}{C_1(t)}$ and $R_2(t)=\frac{C_1(t)}{C_2(t)}$, as discussed in the main text.

\subsection*{S1.2 The winding number of the complex-energy bands}
\begin{figure}
	\centering
	\includegraphics[width=1.0\textwidth]{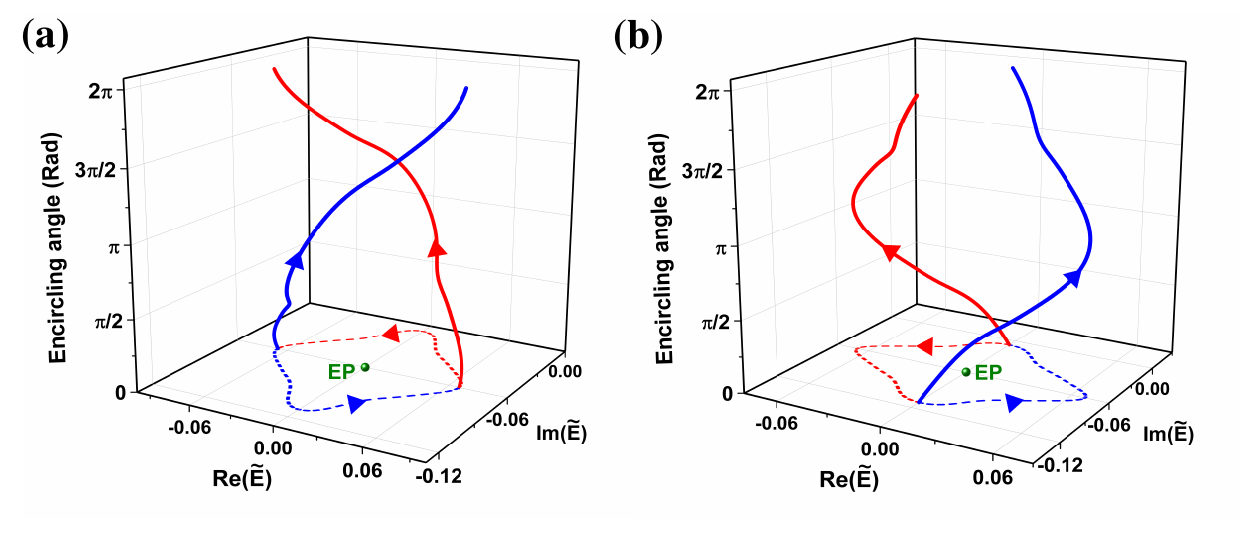}
	\caption{Windings of the complex-energy band structure in parallel transported basis. (a) The encirclement starting from the $\mathcal{PT}$-symmetric regime; (b) The encirclement starting from the $\mathcal{PT}$-broken regime. The encircling angle $\theta \in [0, 2\pi]$ parametrizes the closed loop, and dashed curves are the projection of the energy trajectory. The red and blue colors represent the different bands, and the olive circles represent the EPs.}
	\label{winding}
\end{figure}

The singularity of complex-energy Riemann surface allows for the half-integer windings of the paired bands.Figure 2A in the main text illustrates such a scenario with a winding number $\mathcal{V}=-1/2$. It is important to note that the loop in Fig.~2A does not involve DNAT. To calculate the winding number of a loop that involves DNATs, the evolution processes must be transformed into the parallel transported eigenbasis. In this basis, loops involving DNATs also can be parametrized by the encircling angle $\theta \in [0, 2\pi]$ around the EP.  The theoretical results for $\theta$ in the complex-energy plane are shown in Fig.~\ref{winding}. The solid and dashed curves represent the theoretical
loop and the projection of energy trajectories, respectively. Each band's winding number is 1/2, defining a topological invariant associated with the non-Hermitian band structures~\cite{shen2018topological}. The positive or negative value of the winding number depends on the orientation of encircling curve; it is negative if the curve travels around the EP in a clockwise direction.

\section*{S2~Dynamically encircling of  $\mathcal{APT}$ Hamiltonian}
\subsection*{S2.1 $\mathcal{APT}$ symmetric Hamiltonian}
The construction of the $\mathcal{APT}$-symmetric Hamiltonian has been demonstrated in our previous work~\cite{bian2023quantum}. The evolution operator of the $\mathcal{APT}$ Hamiltonian $U(\tau)=e^{-iH^{\prime}_{eff} t}$, is constructed by sandwiching a passive $\mathcal{PT}$-symmetric Hamiltonian $H_{eff}$ between two $\pi/2$ pulses along the $\pm Y$ axis on the Bloch sphere. First, a $-\pi/2$ microwave pulse is applied along the $Y$ axis in the rotating frame. Then, identical to the construction of the $\mathcal{PT}$-symmetric Hamiltonian, a microwave field along the $X$ axis with strength $J$ and a dissipation laser beam are applied simultaneously for a duration $t$. Finally, a $\pi/2$ pulse along the $Y$ is implemented. The whole evolution is described as
\begin{equation}
	R_y(\frac{\pi}{2})e^{-iH_{eff}t}R_y(-\frac{\pi}{2})=e^{-iR_y(\frac{\pi}{2})H_{eff}R_y(-\frac{\pi}{2})t}
	=e^{-i(2i\gamma I_x-2JI_z-i\gamma \mathbf{I})t}=e^{-i H^{\prime}_{eff}t},
	\label{ry}
\end{equation}
where the passive $\mathcal{APT}$-symmetric Hamiltonian is
\begin{equation}
	H^{\prime}_{eff} = \begin{pmatrix} -J-i\gamma & i\gamma \\ i\gamma & J-i \gamma \end{pmatrix}.
	\label{Hapteff}
\end{equation}
It differs from the $\mathcal{APT}$-symmetric Hamiltonian
\begin{equation}
	H_{APT} = \begin{pmatrix} -J & i\gamma \\ i\gamma & J \end{pmatrix}.
	\label{Hapt}
\end{equation}
by a term $i\gamma$\textbf{I}. $H_{APT}$ satisfies the anti-commutation relation $\{H_\mathcal{APT}, \mathcal{PT}\}=0$ and is a pseudo-Hermitian~\cite{zhang2020synthetic}, satisfying $\sigma_z^{-1}H_\mathcal{APT}\sigma_z = H_\mathcal{APT}^{\dagger}$.
\begin{figure}[h!]
	\includegraphics[width=1.0\linewidth]{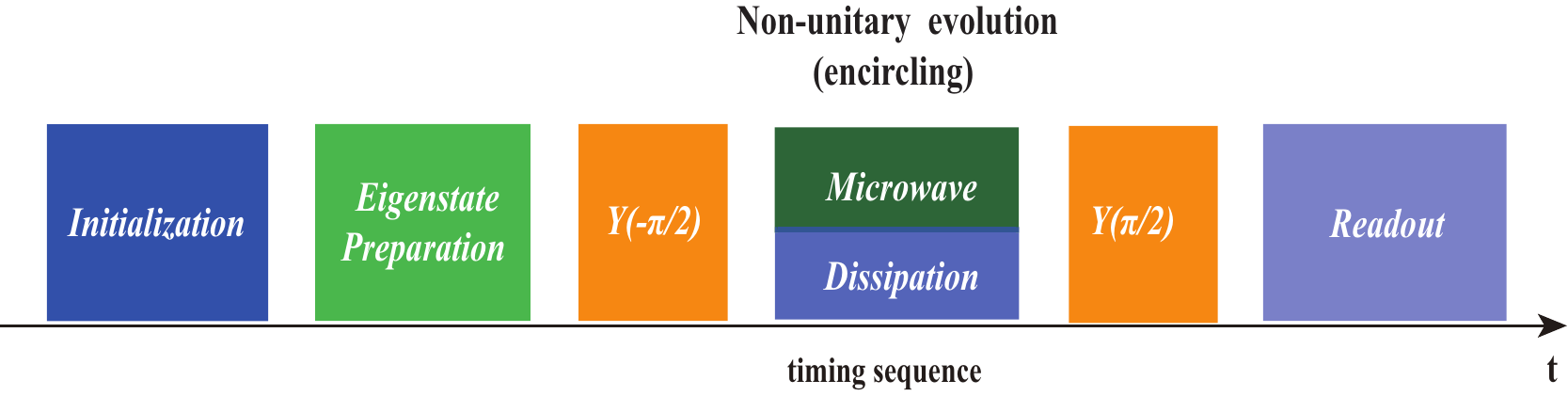}
	\caption{The timing sequence of the coupling microwave and dissipation laser for realizing the dynamical encircling of the EP with the $\mathcal{APT}$-symmetric Hamiltonian. $Y(\pm \pi/2)$ denotes the rotation along the $\pm Y$ axis by $\pi/2$ pulse.}
	\label{apt}
\end{figure}

\subsection*{S2.2 Dynamically encircling the EP of $\mathcal{APT}$ Hamiltonian}
The experimental timing sequence of dynamically encircling the EP of the $\mathcal{APT}$-symmetric Hamiltonian is shown in Fig.~\ref{apt}. Here, the Riemann surfaces defined by $\{\frac{J}{\gamma}, \frac{\Delta}{\gamma}, Re[\lambda]\}$ and $\{\frac{J}{\gamma}, \frac{\Delta}{\gamma}, Im[\lambda]\}$ are identical to those of the $\mathcal{PT}$-symmetric Hamiltonian due to the same complex eigenvalues.

\begin{figure}[h!]
	\centering
	\includegraphics[width=1.0\textwidth]{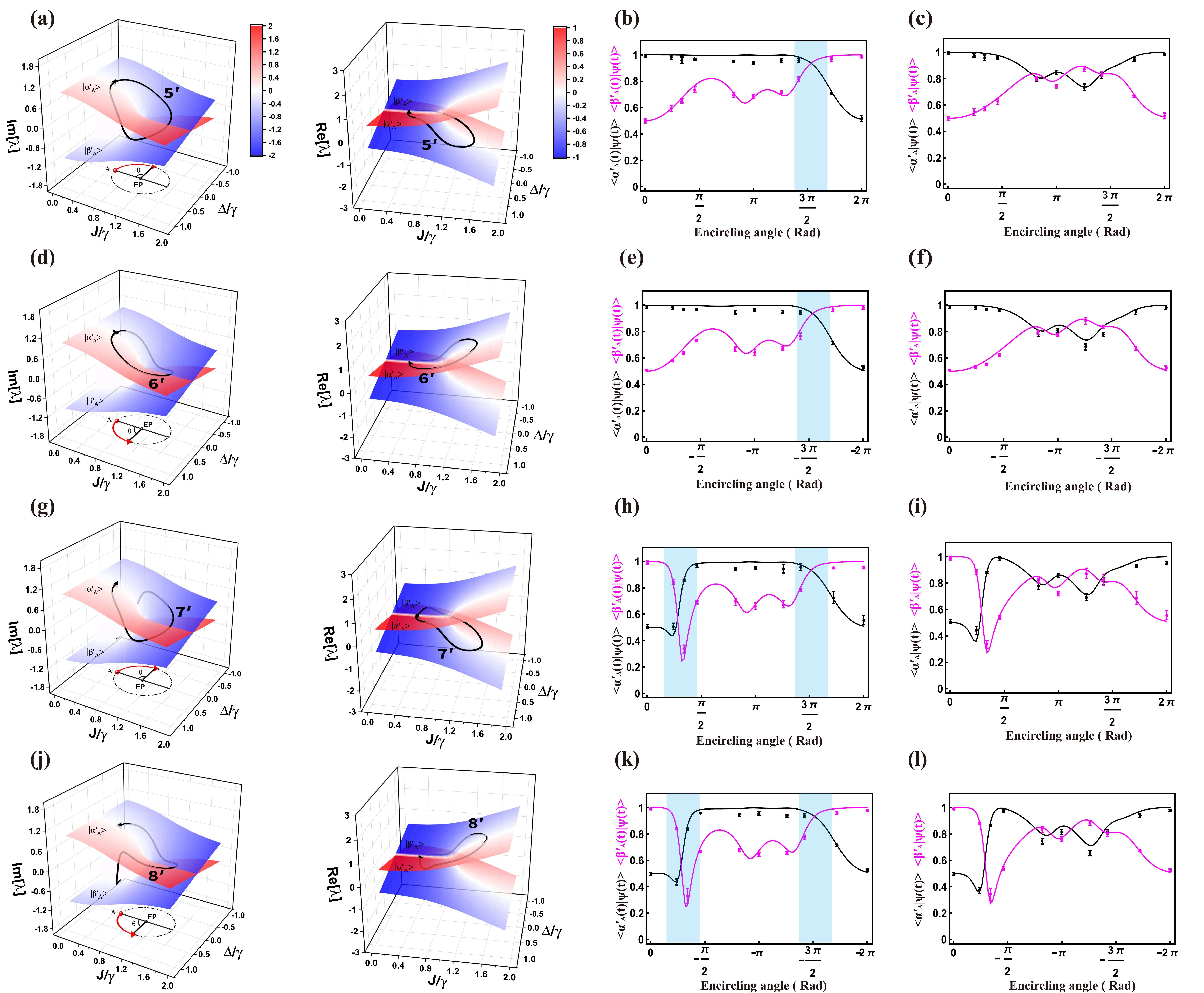}
	\caption{Time evolution of $|\psi(t)\rangle$ starting from the $\mathcal{PTS}$ regime of the $\mathcal{APT}$ symmetric Hamiltonian. The first two columns (a,d,g,j) depict the trajectories of the state evolution on the eigenvalue Riemann surface. The third column (b,e,h,k) depicts the overlap between the evolutionary state and two time-dependent eigenstates $\langle \alpha^{\prime}_{A}(t)|\psi(t)\rangle$ and $\langle \beta^{\prime}_{A}(t)|\psi(t)\rangle$, where the cyan shaded regions indicate the occurrence of  nonadiabatic dynamics. The fourth column (c,f,i,l) represents the overlap between the evolutionary state and two initial eigenstates. (a-c): Clockwise encircling with the initial state $|\alpha^{\prime}_{A}(0)\rangle$. (d-f): Counterclockwise encircling with the initial state $|\alpha^{\prime}_{A}(0)\rangle$. (g-i): Clockwise encircling with the initial state $|\beta^{\prime}_{A}(0)\rangle$. (j-l): Counterclockwise encircling with the initial state $|\beta^{\prime}_{A}(0)\rangle$. Circles and lines represent experimental measurements and numerical simulations, respectively.}
	\label{APTS}
\end{figure}

Fig.~\ref{APTS} shows the state evolution when the initial states of encircling are in the $\mathcal{PTS}$ regime of $\mathcal{APT}$ symmetric Hamiltonian, where the nonreciprocal state transfer can be observed.  For example, the forward-time evolution $|\beta^{\prime}_A(0)\rangle \xrightarrow{\text{CW}} |\alpha^{\prime}_A(0)\rangle$ (trajectory 7$^{\prime}$) and its backward-time evolution $|\alpha^{\prime}_A(0)\rangle \xrightarrow{\text{CCW}} |\alpha^{\prime}_A(0)\rangle$ (trajectory 6$^{\prime}$) illustrate this nonreciprocal behavior. Similarly, the forward-time evolution $|\beta^{\prime}_A(0)\rangle \xrightarrow{\text{CCW}} |\alpha^{\prime}_A(0)\rangle$ (trajectory 8$^{\prime}$) and its backward-time evolution $|\alpha^{\prime}_A(0)\rangle \xrightarrow{\text{CW}} |\alpha^{\prime}_A(0)\rangle$ (trajectory 5$^{\prime}$) further exemplify this nonreciprocal behavior.

\begin{figure}[h!]
	\centering
	\includegraphics[width=1.0\textwidth]{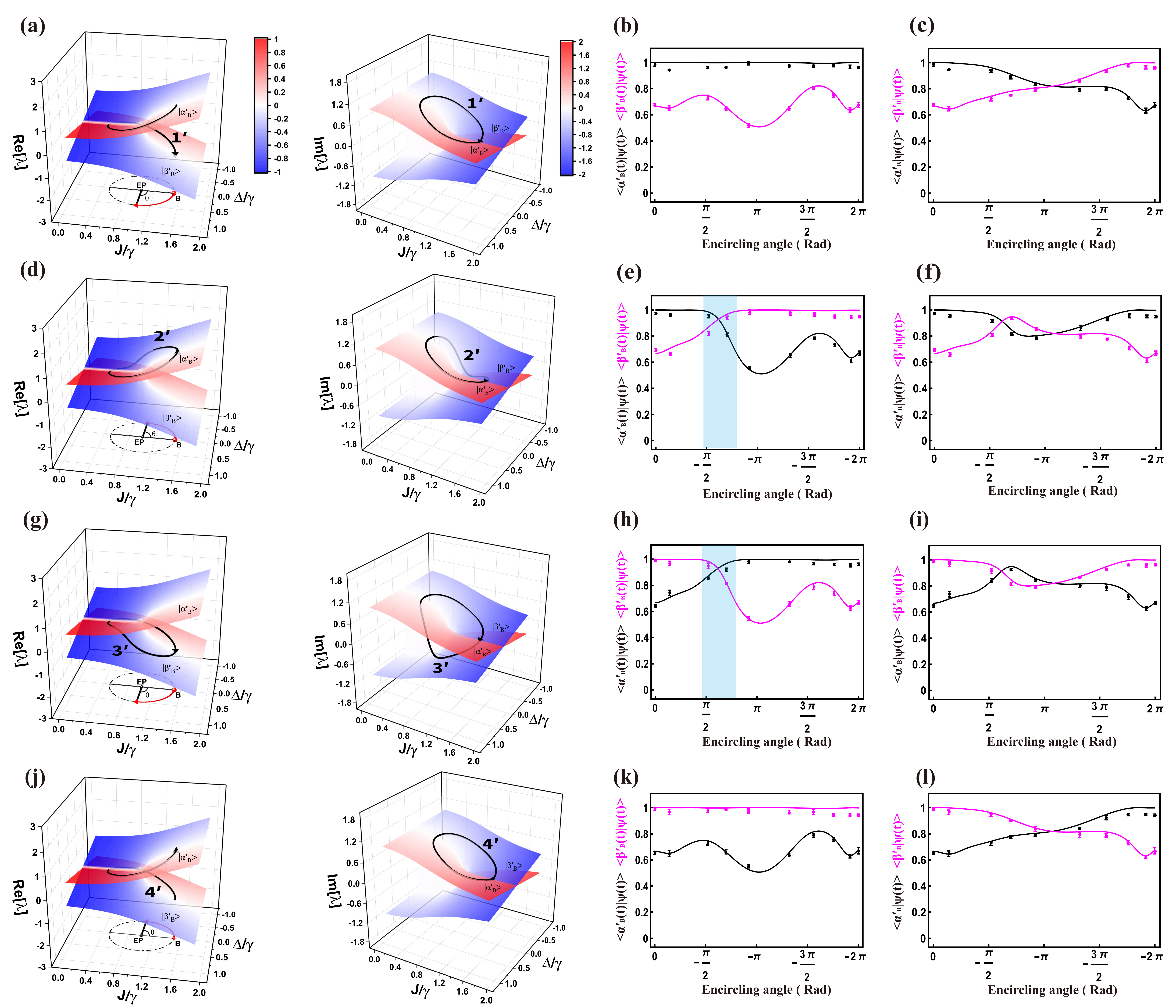}
	\caption{Time evolution of $|\psi(t)\rangle$ starting from the $\mathcal{PTB}$ regime with the  $\mathcal{APT}$ symmetric Hamiltonian. The first two columns (a,d,g,j) depict the trajectories of the state evolution on the eigenvalue Riemann surface. The third column (b,e,h,k) depicts the overlap between the evolutionary state and two time-dependent eigenstates $\langle \alpha^{\prime}_{B}(t)|\psi(t)\rangle$ and $\langle \beta^{\prime}_{B}(t)|\psi(t)\rangle$, where the cyan shaded regions indicate the occurrence of non-adiabatic dynamics. The fourth column (c,f,i,l) represents the overlap between the evolutionary state and two initial eigenstates. (a-c): Clockwise encircling with the initial state $|\alpha^{\prime}_{B}(0)\rangle$. (d-f): Counterclockwise encircling with the initial state  $|\alpha^{\prime}_{B}(0)\rangle$. (g-i): Clockwise encircling with the initial state $|\beta^{\prime}_{B}(0)\rangle$. (j-l): Counterclockwise encircling with the initial state $|\beta^{\prime}_{B}(0)\rangle$. Circles and lines represent experimental measurements and numerical simulations, respectively.}
	\label{APTB}
\end{figure}

Fig.~\ref{APTB} shows the state evolution when the initial state of encircling is in the $\mathcal{PTB}$ regime of the $\mathcal{APT}$ symmetric Hamiltonian, where both the chiral and nonreciprocal state transfer can be observed. The clockwise evolution $|\alpha^{\prime}_B(0)\rangle \xrightarrow{\text{CW}} |\beta^{\prime}_B(0)\rangle$ (trajectory 1$^{\prime}$) and counterclockwise evolution $|\alpha^{\prime}_B(0)\rangle \xrightarrow{\text{CCW}} |\alpha^{\prime}_B(0)\rangle$ (trajectory 2$^{\prime}$) exhibit the chiral behavior, similar to trajectory 3$^{\prime}$ $|\beta^{\prime}_B(0)\rangle \xrightarrow{\text{CW}} |\beta^{\prime}_B(0)\rangle$ and trajectory 4$^{\prime}$ $|\beta^{\prime}_B(0)\rangle \xrightarrow{\text{CCW}} |\alpha^{\prime}_B(0)\rangle$. Nonreciprocal state transfer is also illustruated in the pair of the forward-time evolution $|\alpha^{\prime}_B(0)\rangle \xrightarrow{\text{CCW}} |\alpha^{\prime}_B(0)\rangle$ (trajectory 2$^{\prime}$) and the backward-time evolution $|\alpha^{\prime}_B(0)\rangle \xrightarrow{\text{CW}} |\beta^{\prime}_B(0)\rangle$ (trajectory 1$^{\prime}$), as well as in the pair of trajectory 3$^{\prime}$ $|\beta^{\prime}_B(0)\rangle \xrightarrow{\text{CW}} |\beta^{\prime}_B(0)\rangle$ and trajectory 4$^{\prime}$ $|\beta^{\prime}_B(0)\rangle \xrightarrow{\text{CCW}} |\alpha^{\prime}_B(0)\rangle$.

It is noticed that the chiral and nonreciprocal dynamics of the $\mathcal{PTB}$ regime of the $\mathcal{APT}$ symmetric Hamiltonian are analogous to those in the $\mathcal{PTS}$ regime of the  $\mathcal{PT}$ symmetric Hamiltonian, and vice versa. We summarize the chiral (nonchiral) and reciprocal (nonreciprocal) behaviors of both $\mathcal{PT}$ and $\mathcal{APT}$ symmetric Hamiltonians in Table.~\ref {table1}.

\begin{figure}
	\centering
	\includegraphics[height=1.0\linewidth, width=0.9\linewidth]{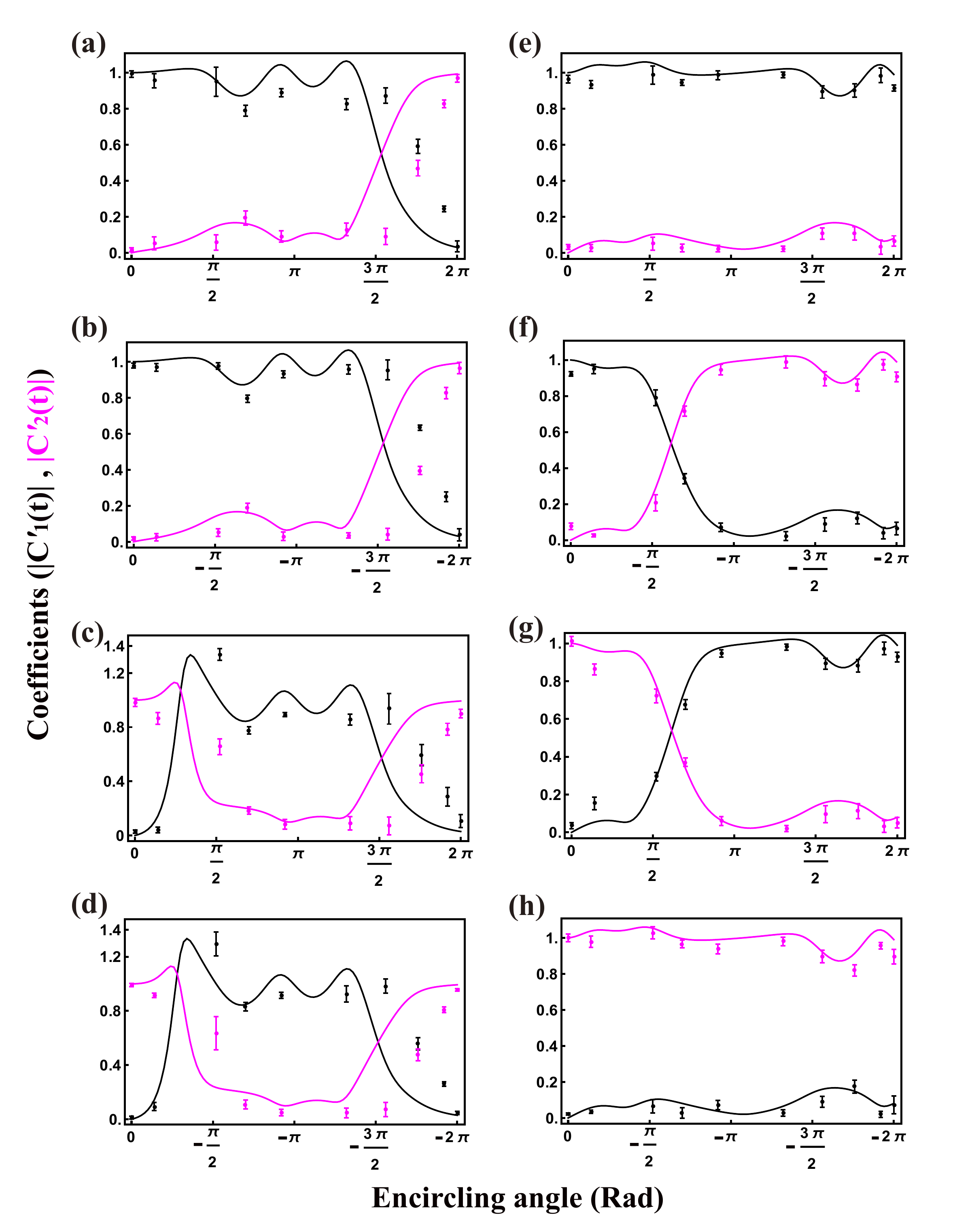}
	\caption{The amplitude coefficients of the instantaneous eigenstates when the $\mathcal{APT}$ Hamiltonian encircles the EP.
		(a-d) Encircling starting from the $\mathcal{PT}$ symmetric regime, with
		(a) Clockwise with the initial state $|\alpha^{\prime}_A(0)\rangle$,
		(b) Counterclockwise with the initial state $|\alpha^{\prime}_A(0)\rangle$,
		(c) Clockwise with the initial state $|\beta^{\prime}_A(0)\rangle$,
		(d) Counterclockwise with the initial state $|\beta^{\prime}_A(0)\rangle$.
		(e-h) Encircling starting from the $\mathcal{PT}$ broken regime, with
		(e) Clockwise with the initial state $|\alpha^{\prime}_B(0)\rangle$,
		(f) Counterclockwise with the initial state $|\alpha^{\prime}_B(0)\rangle$,
		(g) Clockwise with the initial state $|\beta^{\prime}_B(0)\rangle$,
		(h) Counterclockwise with the initial state $|\beta^{\prime}_B(0)\rangle$.
		The amplitude coefficients of $C^{\prime}_1(t)$ and $C^{\prime}_2(t)$ are determined by $|\psi(t)\rangle = C^{\prime}_1(t)|\alpha^{\prime}(t)\rangle + C^{\prime}_2(t)|\beta^{\prime}(t)\rangle$.}
	\label{aptc}
\end{figure}

\begin{figure}
	\centering
	\includegraphics[height=1.0\linewidth, width=0.9\linewidth]{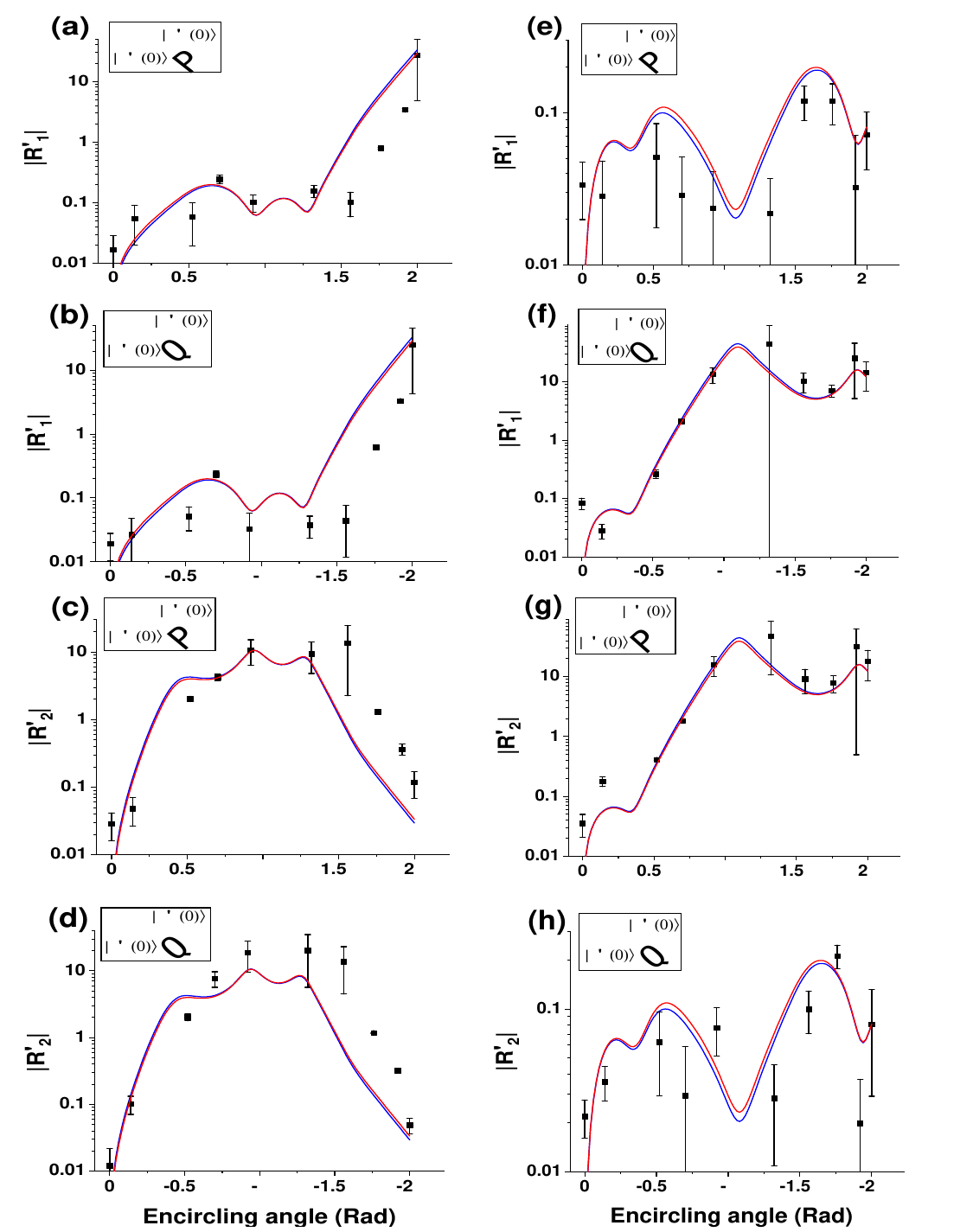}
	\caption{The time-dependent nonadiabatic transition amplitude when the $\mathcal{APT}$ Hamiltonian encircles the EP.
		(a-d) Encircling starting from the $\mathcal{PT}$ symmetric regime, with
		(a) Clockwise with the initial state $|\alpha^{\prime}_A(0)\rangle$,
		(b) Counterclockwise with the initial state $|\alpha^{\prime}_A(0)\rangle$,
		(c) Clockwise with the initial state $|\beta^{\prime}_A(0)\rangle$,
		(d) Counterclockwise with the initial state $|\beta^{\prime}_A(0)\rangle$.
		(e-h) Encircling starting from the $\mathcal{PT}$ broken regime, with
		(e) Clockwise with the initial state $|\alpha^{\prime}_B(0)\rangle$,
		(f) Counterclockwise with the initial state $|\alpha^{\prime}_B(0)\rangle$,
		(g) Clockwise with the initial state $|\beta^{\prime}_B(0)\rangle$,
		(h) Counterclockwise with the initial state $|\beta^{\prime}_B(0)\rangle$.
		The black squares with error bars, red and blue lines correspond to the experimental data, theoretical calculation (Eq.~\ref{rde1}) and numerical simulations from $C^{\prime}_1(t)$ and $C^{\prime}_2(t)$. }
	\label{aptr}
\end{figure}

The experimental results of DNATs for the $\mathcal{APT}$-symmetric Hamiltonian are shown in Fig.~\ref{aptc} and~\ref{aptr}. The notation and procedure are similar to those used in the experiments for measuring DNATs for the $\mathcal{PT}$-symmetric Hamiltonian as described in the main text.

The DNATs with the $\mathcal{APT}$-symmetric Hamiltonian have also been measured in our experiments. In this context, the nonadiabatic transition amplitudes for the $\mathcal{APT}$-symmetric Hamiltonian can be defined using the same method as that employed for the $\mathcal{PT}$-symmetric Hamiltonian.
\begin{equation}
	\begin{split}
		\dot{R'_1}(t) = -2i \lambda R'_1(t) - i f' (1+(R_1'(t))^2)\\
		\dot{R'_2}(t)=  2i \lambda R'_2(t)+ i f' (1+(R_2'(t))^2),
	\end{split}
	\label{rde1}
\end{equation}
where \begin{equation}
	f' = \frac{\dot{J}(i\gamma+\Delta/2)-J(i\dot{\gamma}+\dot{\Delta}/2)}{2i\lambda^2}.
\end{equation}

The time-varying relative nonadiabatic transition amplitude $R^{\prime}_1(t)=\frac{C^{\prime}_2(t)}{C^{\prime}_1(t)}$ and $R^{\prime}_2(t)=\frac{C^{\prime}_1(t)}{C^{\prime}_2(t)}$ can also be obtained by numerically solving Eq.~\ref{rde1}.

\begin{table}
	\renewcommand{\arraystretch}{2.5}	 
	\tabcolsep=2em
	\centering
	\caption{Chiral (nonchiral) and reciprocal (nonreciprocal) dynamic behavior in a dissipative quantum system.}
	\begin{ruledtabular}
		\begin{tabular}{ccccc} 	
			{ } & {Chirality} & {Nonchirality} & {Reciprocity} & {Nonreciprocity}  \\  \hline
			{\makecell[c] {$[\mathcal{PT},H]=0$\\ $\mathcal{PT}|\psi\rangle=|\psi\rangle$}} & {\makecell[c] {Trajectory 1 and 2\\ Trajectory 3 and 4}}& {} & {\makecell[c] {1 and 4 \\ 4 and 1}} &{\makecell[c] { 2 and 1\\  3 and 4}} \\  \hline
			{\makecell[c] {$\{\mathcal{PT},H\}=0$\\ $\mathcal{PT}|\psi\rangle\neq|\psi\rangle$}}  & { \makecell[c]{1$^{\prime}$ and 2$^{\prime}$\\  3$^{\prime}$ and 4$^{\prime}$\\ }} & {} & {\makecell[c] {1$^{\prime}$ and 4$^{\prime}$\\ 4$^{\prime}$ and 1$^{\prime}$\\ }}  & {\makecell[c] { 2$^{\prime}$ and 1$^{\prime}$\\  3$^{\prime}$ and 4$^{\prime}$\\ }}  \\ \hline
			{\makecell[c] {$[\mathcal{PT},H]=0$\\ $\mathcal{PT}|\psi\rangle\neq|\psi\rangle$}}  &{} &{\makecell[c] { 5 and 6\\ 7 and 8\\}} & {\makecell[c] { 5 and 6\\ 6 and 5\\ }} & {\makecell[c] {7 and 6\\  8 and 5\\}}   \\ \hline
			{\makecell[c] {$\{\mathcal{PT},H\}=0$\\ $\mathcal{PT}|\psi\rangle=|\psi\rangle$}}  & {}   & {\makecell[c]{5$^{\prime}$ and 6$^{\prime}$\\  7$^{\prime}$ and 8$^{\prime}$\\ }} & {\makecell[c] { 5$^{\prime}$ and 6$^{\prime}$\\ 6$^{\prime}$ and 5$^{\prime}$\\ }}  &{\makecell[c] {7$^{\prime}$ and 6$^{\prime}$\\ 8$^{\prime}$ and 5$^{\prime}$\\ }}   
		\end{tabular}
	\end{ruledtabular}
	\label{tables1}
\end{table}

\begin{figure}
	\centering
	\includegraphics[height=0.9\linewidth, width=1.0\linewidth]{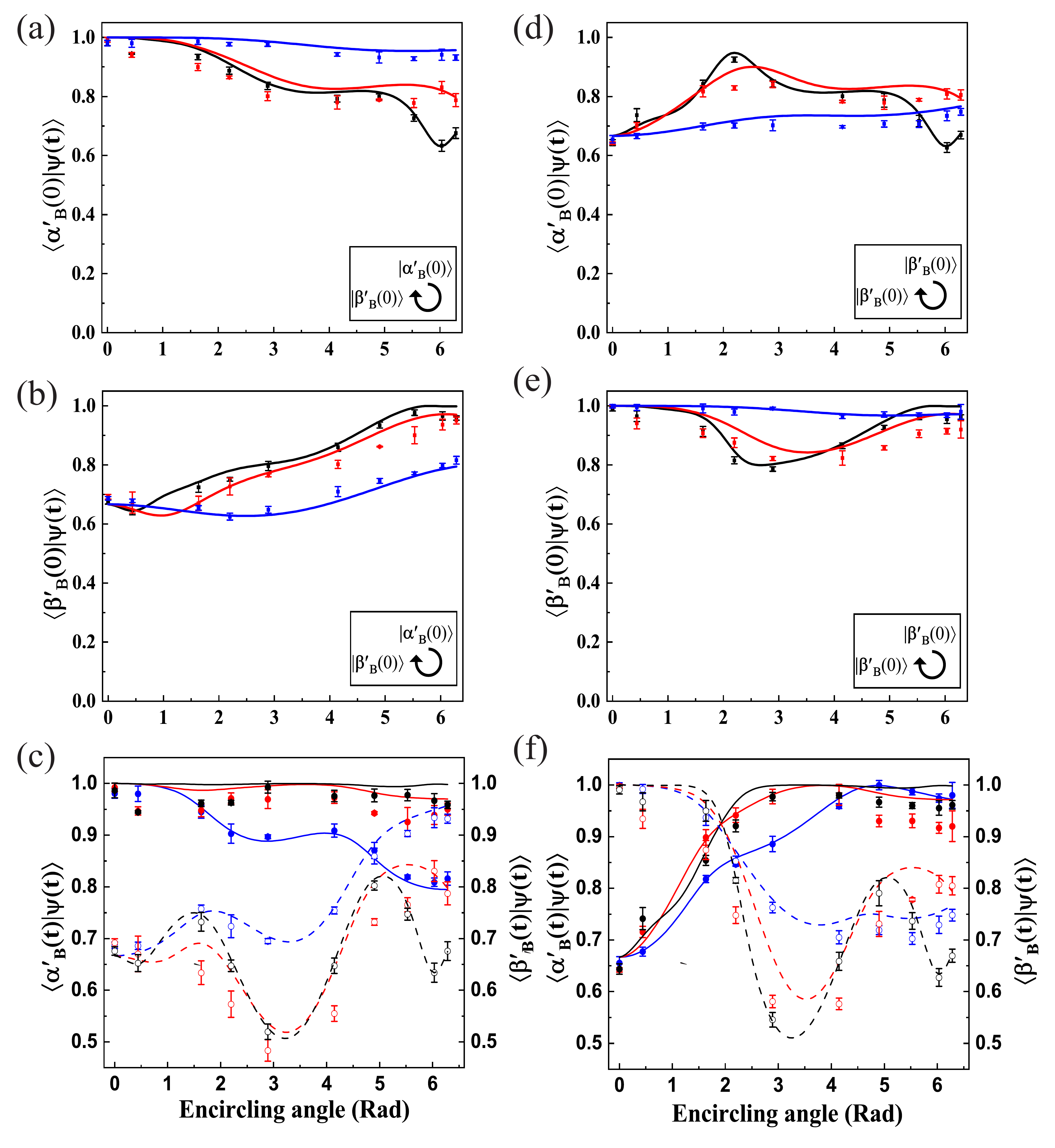}
	\caption{The encircling of the EP of the $\mathcal{APT}$ Hamiltonian with various periods. (a-c) Clockwise encircling the EP with the initial eigenstate $|\alpha^{\prime}_B(0)\rangle$. (a) and (b) are $\langle\alpha^{\prime}_B(0)|\psi(t)\rangle$ and $\langle\beta^{\prime}_{B}(0)|\psi(t)\rangle$, respectively. (c) $\langle\alpha^{\prime}_{B}(t)|\psi(t)\rangle$ (solid line) and $\langle\beta^{\prime}_{B}(t)|\psi(t)\rangle$ (dash line).
		(d-f) Clockwise encircling the EP with the initial eigenstate $|\beta^{\prime}_B(0)\rangle$, where the notation in (d-f) are the same as (a-c). The blue, red, and black circles correspond to $T=16.67~\mu s$, $100~\mu s$, and $250~\mu s$, respectively. The lines represent the numerical simulations.}
	\label{aptt}
\end{figure}

\begin{figure}
	\centering
	\includegraphics[height=0.9\linewidth, width=0.9\linewidth]{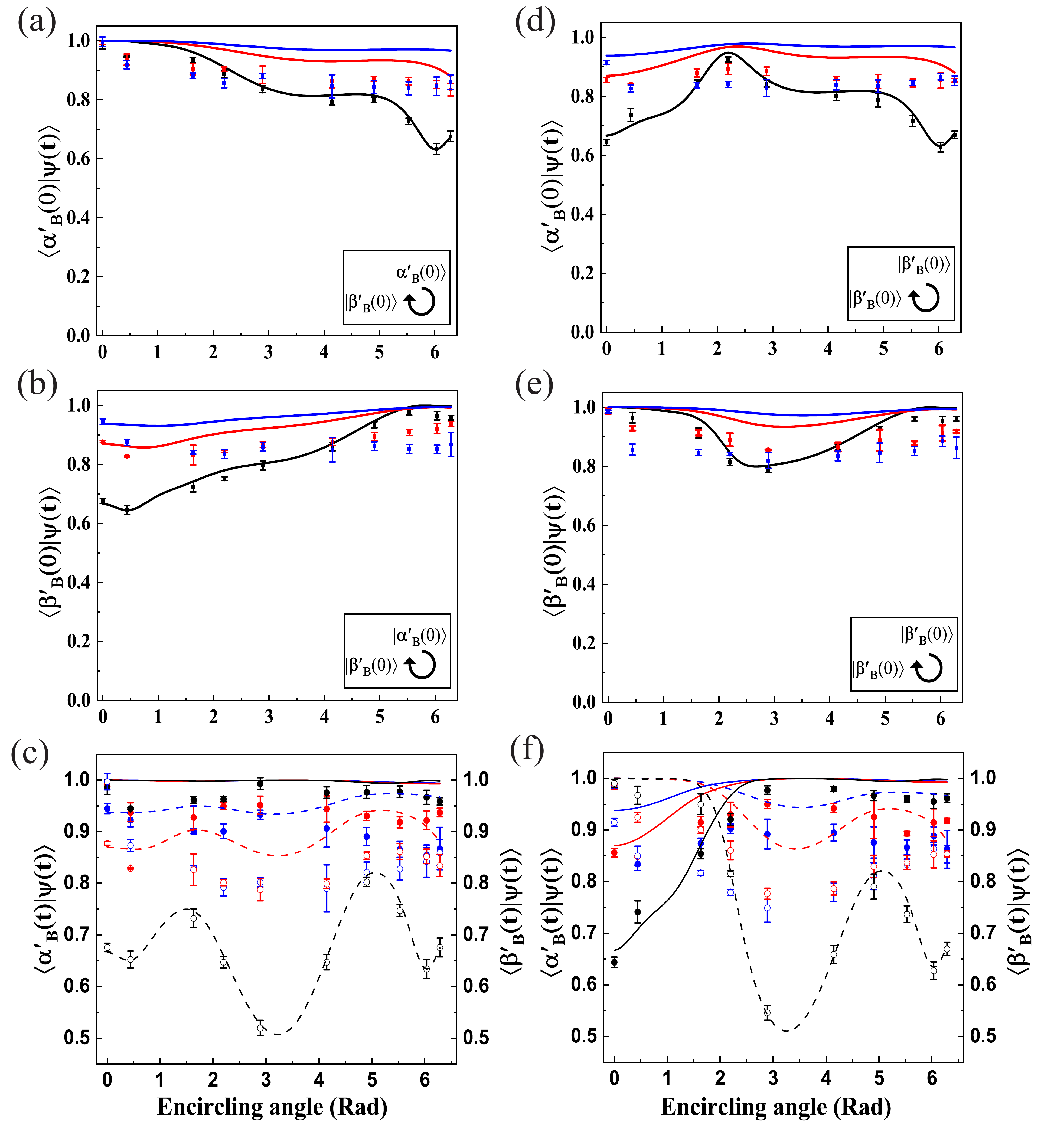}
	\caption{The encircling of the EP of the $\mathcal{APT}$ Hamiltonian with various radii. The notation for (a-f) is similar to Fig.~\ref{aptt} (a-f) as defined previously. The blue, red, and black squares correspond to $r=$0.004 MHz, 0.009 MHz, and 0.03 MHz, respectively. The lines represent the numerical simulations.}
	\label{aptrr}
\end{figure}

Furthermore, we investigate the adiabatic theorem for dynamical encircling of the EP with the  $\mathcal{APT}$-symmetric Hamiltonian. The details are presented in the Fig.~\ref{aptt} and Fig.~\ref{aptrr}. These observations are quite similar to those with the $\mathcal{PT}$-symmetric Hamiltonian, as expected since they share the same Riemann surface of the eigenvalues.

\subsection*{S2.3 Topological robustness of quantum state transfer with $\mathcal{APT}$ Hamiltonian}
\begin{figure}
	\centering
	\includegraphics[width=1.0\textwidth]{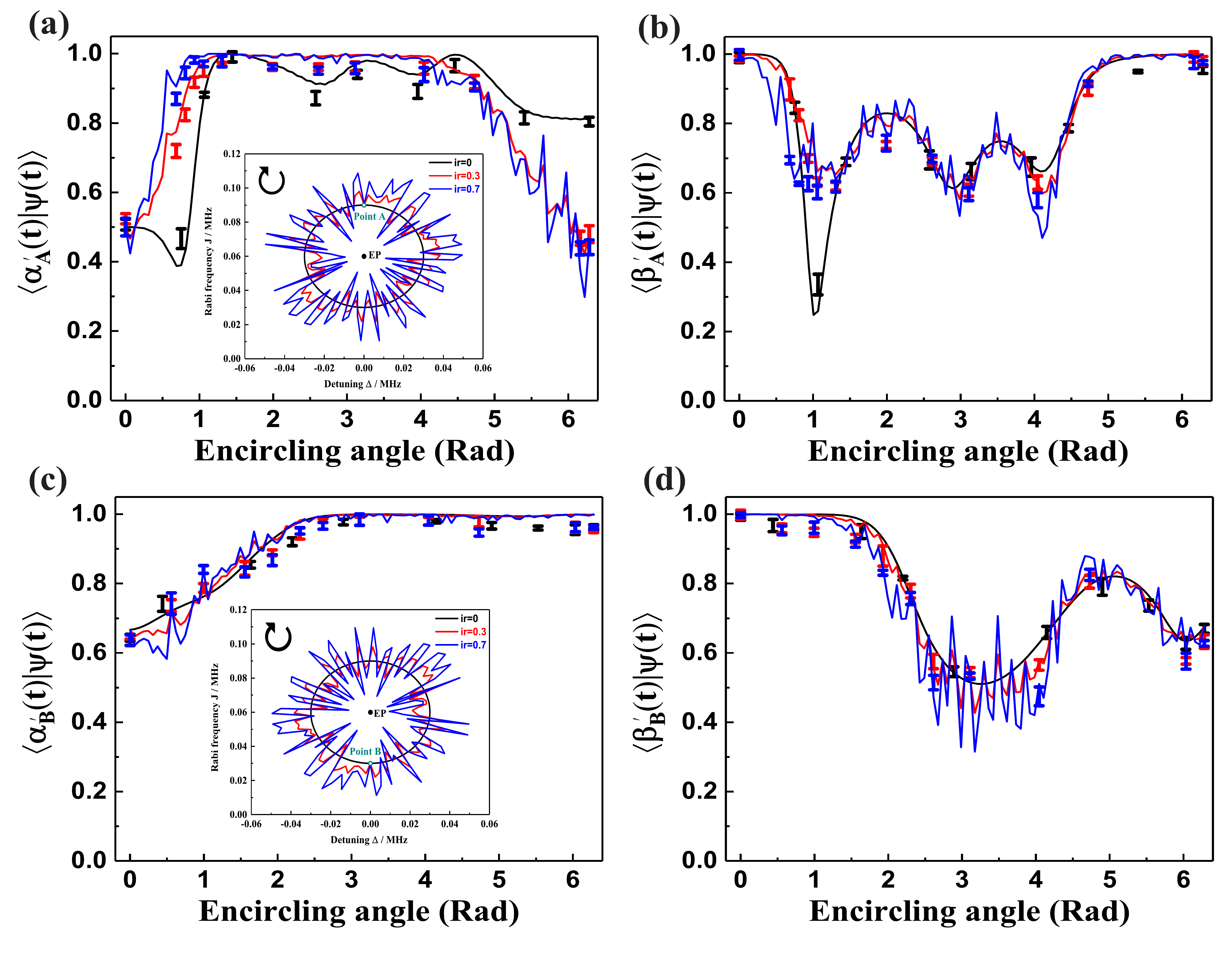}
	\caption{Quantum state transfers while encircling the EP of the $\mathcal{APT}$ Hamiltonian with various noise intensities. (a-b) The clockwise encircling from the initial state $|\beta^{\prime}_A(0)\rangle$. (c-d) The clockwise encircling from the initial state $|\beta^{\prime}_B(0)\rangle$. The insets in (a) and (c) display the encircling trajectories in the parameter space $\{\Delta, J\}$ with random noise intensities $ir =$ 0 (black), 0.3 (red), and 0.7 (blue), respectively. The dots are experimental results, while the solid lines represent the numerical simulations.}
	\label{APTN}
\end{figure}

The results with the initial states of $|\beta^{\prime}_A(0)\rangle$ (in the $\mathcal{PTS}$ regime) and $|\beta^{\prime}_B(0)\rangle$ (in the $\mathcal{PTB}$ regime) are shown in Fig.~\ref{APTN}(a-b) and Fig.~\ref{APTN}(c-d), respectively. In both cases, the experimental data match well with the theoretically
predicted values despite varying noise intensities, thereby validating the robustness of topological state transfers with the $\mathcal{APT}$-symmetric Hamiltonian.


\begin{thebibliography}{41}
	
\bibitem{schnyder2008classification}
	A.~P. Schnyder, S.~Ryu, A.~Furusaki, A.~W. Ludwig, Classification of
	topological insulators and superconductors in three spatial dimensions.
	\newblock {\it Physical Review B\/} {\bf 78}, 195125 (2008).
	
\bibitem{chiu2016classification}
	C.-K. Chiu, J.~C. Teo, A.~P. Schnyder, S.~Ryu, Classification of topological
	quantum matter with symmetries.
	\newblock {\it Reviews of Modern Physics\/} {\bf 88}, 035005 (2016).
	
\bibitem{cohen2019geometric}
	E.~Cohen, H.~Larocque, F.~Bouchard, F.~Nejadsattari, Y.~Gefen, E.~Karimi,
	Geometric phase from aharonov--bohm to pancharatnam--berry and beyond.
	\newblock {\it Nature Reviews Physics\/} {\bf 1}, 437--449 (2019).
	
\bibitem{zhang2005experimental}
	Y.~Zhang, Y.-W. Tan, H.~L. Stormer, P.~Kim, Experimental observation of the
	quantum hall effect and berry's phase in graphene.
	\newblock {\it Nature\/} {\bf 438}, 201--204 (2005).
	
\bibitem{hasan2010colloquium}
	M.~Z. Hasan, C.~L. Kane, Colloquium: topological insulators.
	\newblock {\it Reviews of Modern Physics\/} {\bf 82}, 3045 (2010).
	
\bibitem{xu2011topological}
	S.-Y. Xu, Y.~Xia, L.~Wray, S.~Jia, F.~Meier, J.~Dil, J.~Osterwalder,
	B.~Slomski, A.~Bansil, H.~Lin, {\it et~al.\/}, Topological phase transition
	and texture inversion in a tunable topological insulator.
	\newblock {\it Science\/} {\bf 332}, 560--564 (2011).
	
\bibitem{doppler2016dynamically}
	J.~Doppler, A.~A. Mailybaev, J.~B{\"o}hm, U.~Kuhl, A.~Girschik, F.~Libisch,
	T.~J. Milburn, P.~Rabl, N.~Moiseyev, S.~Rotter, Dynamically encircling an
	exceptional point for asymmetric mode switching.
	\newblock {\it Nature\/} {\bf 537}, 76--79 (2016).
	
\bibitem{xu2016topological}
	H.~Xu, D.~Mason, L.~Jiang, J.~Harris, Topological energy transfer in an
	optomechanical system with exceptional points.
	\newblock {\it Nature\/} {\bf 537}, 80--83 (2016).
	
\bibitem{dembowski2001experimental}
	C.~Dembowski, H.-D. Gr{\"a}f, H.~Harney, A.~Heine, W.~Heiss, H.~Rehfeld,
	A.~Richter, Experimental observation of the topological structure of
	exceptional points.
	\newblock {\it Physical Review Letters\/} {\bf 86}, 787 (2001).
	
\bibitem{zhang2018dynamically}
	X.-L. Zhang, S.~Wang, B.~Hou, C.~T. Chan, Dynamically encircling exceptional
	points: in situ control of encircling loops and the role of the starting
	point.
	\newblock {\it Physical Review X\/} {\bf 8}, 021066 (2018).
	
\bibitem{yoon2018time}
	J.~W. Yoon, Y.~Choi, C.~Hahn, G.~Kim, S.~H. Song, K.-Y. Yang, J.~Y. Lee,
	Y.~Kim, C.~S. Lee, J.~K. Shin, {\it et~al.\/}, Time-asymmetric loop around an
	exceptional point over the full optical communications band.
	\newblock {\it Nature\/} {\bf 562}, 86--90 (2018).
	
\bibitem{li2020hamiltonian}
	A.~Li, J.~Dong, J.~Wang, Z.~Cheng, J.~S. Ho, D.~Zhang, J.~Wen, X.-L. Zhang,
	C.~T. Chan, A.~Al{\`u}, {\it et~al.\/}, Hamiltonian hopping for efficient
	chiral mode switching in encircling exceptional points.
	\newblock {\it Physical Review Letters\/} {\bf 125}, 187403 (2020).
	
\bibitem{nasari2022observation}
	H.~Nasari, G.~Lopez-Galmiche, H.~E. Lopez-Aviles, A.~Schumer, A.~U. Hassan,
	Q.~Zhong, S.~Rotter, P.~LiKamWa, D.~N. Christodoulides, M.~Khajavikhan,
	Observation of chiral state transfer without encircling an exceptional point.
	\newblock {\it Nature\/} {\bf 605}, 256--261 (2022).
	
\bibitem{yang2023realization}
	M.~Yang, H.-Q. Zhang, Y.-W. Liao, Z.-H. Liu, Z.-W. Zhou, X.-X. Zhou, J.-S. Xu,
	Y.-J. Han, C.-F. Li, G.-C. Guo, Realization of exceptional points along a
	synthetic orbital angular momentum dimension.
	\newblock {\it Science Advances\/} {\bf 9}, eabp8943 (2023).
	
\bibitem{ding2016emergence}
	K.~Ding, G.~Ma, M.~Xiao, Z.~Zhang, C.~T. Chan, Emergence, coalescence, and
	topological properties of multiple exceptional points and their experimental
	realization.
	\newblock {\it Physical Review X\/} {\bf 6}, 021007 (2016).

\bibitem{tang2020exceptional}
	W.~Tang, X.~Jiang, K.~Ding, Y.-X. Xiao, Z.-Q. Zhang, C.~T. Chan, G.~Ma,
	Exceptional nexus with a hybrid topological invariant.
	\newblock {\it Science\/} {\bf 370}, 1077--1080 (2020).
	
\bibitem{abbasi2022topological}
	M.~Abbasi, W.~Chen, M.~Naghiloo, Y.~N. Joglekar, K.~W. Murch, Topological
	quantum state control through exceptional-point proximity.
	\newblock {\it Physical Review Letters\/} {\bf 128}, 160401 (2022).
	
\bibitem{liu2021dynamically}
	W.~Liu, Y.~Wu, C.-K. Duan, X.~Rong, J.~Du, Dynamically encircling an
	exceptional point in a real quantum system.
	\newblock {\it Physical Review Letters\/} {\bf 126}, 170506 (2021).
	
\bibitem{ren2022chiral}
	Z.~Ren, D.~Liu, E.~Zhao, C.~He, K.~K. Pak, J.~Li, G.-B. Jo, Chiral control of
	quantum states in non-hermitian spin--orbit-coupled fermions.
	\newblock {\it Nature Physics\/} {\bf 18}, 385--389 (2022).
	
\bibitem{shen2018topological}
	H.~Shen, B.~Zhen, L.~Fu, Topological band theory for non-hermitian
	hamiltonians.
	\newblock {\it Physical Review Letters\/} {\bf 120}, 146402 (2018).
		
\bibitem{kawabata2018anomalous}
	K.~Kawabata, K.~Shiozaki, M.~Ueda, Anomalous helical edge states in a
	non-hermitian chern insulator.
	\newblock {\it Physical Review B\/} {\bf 98}, 165148 (2018).
	
\bibitem{lu2024realizing}
	P.~Lu, T.~Liu, Y.~Liu, X.~Rao, Q.~Lao, H.~Wu, F.~Zhu, L.~Luo, Realizing quantum
	speed limit in open system with a-symmetric trapped-ion qubit.
	\newblock {\it New Journal of Physics\/} {\bf 26}, 013043 (2024).
	
\bibitem{bian2023quantum}
	J.~Bian, P.~Lu, T.~Liu, H.~Wu, X.~Rao, K.~Wang, Q.~Lao, Y.~Liu, F.~Zhu, L.~Luo,
	Quantum simulation of a general anti-pt-symmetric hamiltonian with a trapped
	ion qubit.
	\newblock {\it Fundamental Research\/} {\bf 3}, 904--908 (2023).
	
\bibitem{bian2022PhRvA}
	J.~Bian, K.~Wang, P.~Lu, X.~Rao, H.~Wu, Q.~Lao, T.~Liu, Y.~Liu, F.~Zhu, L.~Luo,
	Protection of quantum evolutions under parity-time-symmetric non-hermitian
	hamiltonians by dynamical decoupling.
	\newblock {\it Physical Review A\/} {\bf 106}, 012416 (2022).
	
\bibitem{li2019observation}
	J.~Li, A.~K. Harter, J.~Liu, L.~de~Melo, Y.~N. Joglekar, L.~Luo, Observation of
	parity-time symmetry breaking transitions in a dissipative floquet system of
	ultracold atoms.
	\newblock {\it Nature Communications\/} {\bf 10}, 855 (2019).
	
\bibitem{naghiloo2019quantum}
	M.~Naghiloo, M.~Abbasi, Y.~N. Joglekar, K.~Murch, Quantum state tomography
	across the exceptional point in a single dissipative qubit.
	\newblock {\it Nature Physics\/} {\bf 15}, 1232--1236 (2019).
	
\bibitem{garrison1988complex}
	J.~Garrison, E.~Wright, Complex geometrical phases for dissipative systems.
	\newblock {\it Physics Letters A\/} {\bf 128}, 177--181 (1988).
	
\bibitem{milburn2015general}
	T.~J. Milburn, J.~Doppler, C.~A. Holmes, S.~Portolan, S.~Rotter, P.~Rabl,
	General description of quasiadiabatic dynamical phenomena near exceptional
	points.
	\newblock {\it Physical Review A\/} {\bf 92}, 052124 (2015).
	
\bibitem{gao2015observation}
	T.~Gao, E.~Estrecho, K.~Bliokh, T.~Liew, M.~Fraser, S.~Brodbeck, M.~Kamp,
	C.~Schneider, S.~H{\"o}fling, Y.~Yamamoto, {\it et~al.\/}, Observation of
	non-hermitian degeneracies in a chaotic exciton-polariton billiard.
	\newblock {\it Nature\/} {\bf 526}, 554--558 (2015).
	
\bibitem{zhou2018observation}
	H.~Zhou, C.~Peng, Y.~Yoon, C.~W. Hsu, K.~A. Nelson, L.~Fu, J.~D. Joannopoulos,
	M.~Solja{\v{c}}i{\'c}, B.~Zhen, Observation of bulk fermi arc and
	polarization half charge from paired exceptional points.
	\newblock {\it Science\/} {\bf 359}, 1009--1012 (2018).
	
\bibitem{su2021direct}
	R.~Su, E.~Estrecho, D.~Biega{\'n}ska, Y.~Huang, M.~Wurdack, M.~Pieczarka, A.~G.
	Truscott, T.~C. Liew, E.~A. Ostrovskaya, Q.~Xiong, Direct measurement of a
	non-hermitian topological invariant in a hybrid light-matter system.
	\newblock {\it Science Advances\/} {\bf 7}, eabj8905 (2021).
	
\bibitem{ding2022non}
	K.~Ding, C.~Fang, G.~Ma, Non-hermitian topology and exceptional-point
	geometries.
	\newblock {\it Nature Reviews Physics\/} {\bf 4}, 745--760 (2022).
	
\bibitem{berry1984quantal}
	M.~V. Berry, Quantal phase factors accompanying adiabatic changes.
	\newblock {\it Proceedings of the Royal Society of London. A. Mathematical and
		Physical Sciences\/} {\bf 392}, 45--57 (1984).
	
\bibitem{qi2018defect}
	B.~Qi, L.~Zhang, L.~Ge, Defect states emerging from a non-hermitian flatband of
	photonic zero modes.
	\newblock {\it Physical Review Letters\/} {\bf 120}, 093901 (2018).
	
\bibitem{bender2002complex}
	C.~M. Bender, D.~C. Brody, H.~F. Jones, Complex extension of quantum mechanics.
	\newblock {\it Physical Review Letters\/} {\bf 89}, 270401 (2002).
	
\bibitem{wang2018non}
	H.~Wang, L.-J. Lang, Y.~D. Chong, Non-hermitian dynamics of slowly varying
	hamiltonians.
	\newblock {\it Physical Review A\/} {\bf 98}, 012119 (2018).
	
\bibitem{berry2011optical}
	M.~Berry, Optical polarization evolution near a non-hermitian degeneracy.
	\newblock {\it Journal of Optics\/} {\bf 13}, 115701 (2011).
	
\bibitem{ding2021experimental}
	L.~Ding, K.~Shi, Q.~Zhang, D.~Shen, X.~Zhang, W.~Zhang, Experimental
	determination of pt-symmetric exceptional points in a single trapped ion.
	\newblock {\it Physical Review Letters\/} {\bf 126}, 083604 (2021).
	
\bibitem{olde1995stokes}
	A.~Olde~Daalhuis, S.~J. Chapman, J.~R. King, J.~R. Ockendon, R.~H. Tew, Stokes
	phenomenon and matched asymptotic expansions.
	\newblock {\it SIAM Journal on Applied Mathematics\/} {\bf 55}, 1469--1483
	(1995).
	
\bibitem{berry2011slow}
	M.~Berry, R.~Uzdin, Slow non-hermitian cycling: exact solutions and the stokes
	phenomenon.
	\newblock {\it Journal of Physics A: Mathematical and Theoretical\/} {\bf 44},
	435303 (2011).
	
\bibitem{feilhauer2020encircling}
	J.~Feilhauer, A.~Schumer, J.~Doppler, A.~A. Mailybaev, J.~B{\"o}hm, U.~Kuhl,
	N.~Moiseyev, S.~Rotter, Encircling exceptional points as a non-hermitian
	extension of rapid adiabatic passage.
	\newblock {\it Physical Review A\/} {\bf 102}, 040201 (2020).
	
\end{thebibliography}
\end{document}